\documentclass[10pt]{article}
\pdfoutput=1
\usepackage{graphics,color}
\usepackage{bm}
\usepackage{graphicx}
\usepackage{amsmath}
\usepackage{amssymb}
\usepackage{subfigure}
\usepackage{authblk}
\usepackage[margin=0.68in,footskip=0.2in]{geometry}
\usepackage{tikz}
\usepackage{enumitem}

\def\be{\begin{equation}}
\def\ee{\end{equation}}
\def\ba{\begin{eqnarray}}
\def\ea{\end{eqnarray}}
\def\l{\left}
\def\r{\right}
\def\f{\frac}

\title{Effective Field Theory of Dark Energy: a Dynamical Analysis}

\author[1,2]{Noemi Frusciante\thanks{E-mail:\textit{nfruscia@sissa.it}}}
\author[1]{Marco Raveri\thanks{E-mail:\textit{mraveri@sissa.it}}}
\author[1,2,3]{Alessandra Silvestri\thanks{E-mail:\textit{asilvest@sissa.it}}}
\affil[1]{SISSA - International School for Advanced Studies, Via Bonomea 265, 34136, Trieste, Italy}
\affil[2]{INFN, Sezione di Trieste, Via Valerio 2, I-34127 Trieste, Italy }
\affil[3]{INAF-Osservatorio Astronomico di Trieste, Via G.B. Tiepolo 11, I-34131 Trieste, Italy}

\begin{document}
\maketitle

\begin{abstract}
The effective field theory (EFT) of dark energy relies on three functions of time to describe the dynamics of background cosmology. The viability of these functions is investigated here by means of a thorough dynamical analysis. While the system is underdetermined, and one can always find a set of functions reproducing any expansion history, we are able to determine general compatibility conditions for these functions by requiring a viable background cosmology. In particular, we identify a set of variables that allows us to transform the non-autonomous system of equations into an infinite-dimensional one characterized by a significant recursive structure. We then analyze several autonomous sub-systems, obtained truncating the original one at increasingly higher dimension, that correspond to increasingly general models of dark energy and modified gravity. Furthermore, we exploit the recursive nature of the system to draw some general conclusions on the different cosmologies that can be recovered within the EFT formalism and the corresponding compatibility requirements for the EFT functions. 
The machinery that we set up serves different purposes. It offers a general scheme for performing dynamical analysis of dark energy and modified gravity models within the model independent framework of EFT; the general results, obtained with this technique, can be projected into specific models, as we show in one example. It also can be used to determine appropriate ans$\ddot{\text{a}}$tze for the three EFT background functions when studying the dynamics of cosmological perturbations in the context of large scale structure tests of gravity.
\end{abstract}

\maketitle
\newpage

\tableofcontents

\section{Introduction} \label{Sec1}
 Modern Cosmology faces some major challenges, one of which is the phase of accelerated expansion experienced by the late time Universe~\cite{Riess:1998cb,Perlmutter:1998np}, commonly referred to as cosmic acceleration. A universe described by General Relativity and filled with matter, is naturally expected to decelerate after the initial phase of rapid expansion following the big bang. The observed acceleration on large scales requires either an additional component, such as a static or dynamical dark energy, or a modification of the laws of gravity on the largest scales, commonly referred to as modified gravity.
Given the plethora of models that address the phenomenon of cosmic acceleration~\cite{Silvestri:2009hh}, and given the wealth of high precision large scale data  that will come readily available via upcoming and future experiments like Dark Energy Survey~\cite{des}, EUCLID~\cite{euclid,Amendola:2012ys} and Large Synoptic Survey Telescope~\cite{lsst}, it is crucial and timely to identify the optimal parametrization that will allow model independent tests of gravity. In the recent years, there has been a wide effort in the community in this direction, with several alternatives being put forward and analyzed~\cite{Linder:2007hg}-~\cite{Silvestri:2013ne}. 

In the quest for the optimal framework to perform cosmological tests of gravity, some authors have recently proposed an `effective' approach to unify dark energy and modified gravity modeling~\cite{Creminelli:2008wc}-\cite{Bloomfield:2012ff}, inspired by effective field theories of inflation~\cite{Cheung:2007st,Weinberg:2008hq} and large scale structure~\cite{Carrasco:2012cv}. We will refer to it as the \emph{effective field theory of dark energy} (hereafter EFT). The aim of this approach, in particular the one in~\cite{Gubitosi:2012hu,Bloomfield:2012ff}, is that of creating a model independent framework that encompasses all single-field dark energy and modified gravity models, describing the evolution of the background cosmology and of perturbations with a finite number of functions of time introduced at the level of the action. The action is written in unitary gauge, in terms of an expansion in the operators that are compatible with the residual symmetries of unbroken spatial diffeomorphisms, organized in powers of the number of perturbations. There is a finite  number of such operators for each order of perturbation; in particular, only three functions affect the background dynamics. In principle it is possible to link each operator with the corresponding observable effects it produces on the cosmology~\cite{Gubitosi:2012hu,Bloomfield:2012ff}, although in practice such an identification is not always feasible. 

This formalism has been conjectured as a unified description of dark energy and modified gravity to apply to tests involving data on linear cosmological perturbations, therefore it is generally assumed that a background evolution will be specified {\it a priori}; in other words, the background functions will be chosen to closely mimic the evolution of the standard cosmological model ($\Lambda$CDM), and one focuses on effects at the level of perturbations. However, fixing the expansion history does not determine all the EFT functions, and there remain one completely free function of time out of the original three. Given this high degree of freedom, and given that the remaining function affects also the evolution of perturbations, it is important to explore what general viability/compatibility rules can be placed on the background EFT functions by requiring that the corresponding model gives a viable expansion history, rather than fixing the latter a priori. Besides allowing us to exploit the valuable information from geometrical probes, these conditions will guide us in choosing appropriate ans$\ddot{\text{a}}$tze for these functions when moving to the main goal of the the EFT approach, that is studying the evolution of linear cosmological perturbations in a model independent way. 

In this paper we perform a dynamical analysis of the background cosmology, treating the three EFT functions as unknown functions of time. Despite the underdetermined nature of the system, we identify a set of variables that allows us to write it as an infinite-dimensional system with an important recursive structure. We then analyze several autonomous subsystems  of increasingly higher dimension that, as we will illustrate, correspond to higher differential order for combinations of the EFT functions; in other words, we explore more and more general  models of dark energy/modified gravity identifying at each order conditions of cosmological viability. Furthermore, we exploit the recursive nature of the system to draw quite general conclusions on its cosmological dynamics, building on our findings at the lower orders. While we apply our method to some specific cases in order to elucidate it, the machinery we set up is general and can be used to perform dynamical analysis of models of dark energy/modified gravity within the broad and model independent EFT framework.

The paper is organized as follows. In Sec.~\ref{Sec2} we briefly review the EFT formalism, focusing on the terms and equations that are of interest for the background cosmology. In Sec.~\ref{Sec3} we set up the dynamical system, describe our strategy to make it autonomous and then proceed with the dynamical analysis at different, increasing, orders in Sec.~\ref{Seczero},~\ref{Secfirst} and~\ref{Secsecond}, investigating the cosmology of selected trajectories. In Sec.~\ref{recursive} we exploit the recursive nature of our system of equations, as well as the results from the previous analyses, to perform the dynamical analysis at a generic order $N$. Finally we discuss our results and conclude in Sec.~\ref{conclusions}.

\section{Effective Field Theory of Dark Energy}
\label{Sec2} 
We follow the work of~\cite{Gubitosi:2012hu,Bloomfield:2012ff} and consider the effective field theory of dark energy in Jordan frame, described by the following action in unitary gauge:
\begin{align}
\label{eq:theaction}
  S = \int d^4x \sqrt{-g} {}& \left[ \frac{m_0^2}{2} \Omega(t) R + \Lambda(t) - c(t) \delta g^{00}\right] + S^{(2)}_{\rm DE} +  S_{m} [g_{\mu \nu}],
\end{align}
where  $m_0^2$ is the bare Planck mass, $R$ the Ricci scalar, $\delta g^{00}=g^{00}+1$ the perturbation to the upper time-time component of the metric and $\Omega(t)$, $\Lambda(t)$ and $c(t)$ are free functions of the time coordinate $t$. Our notation follows more closely that of~\cite{Bloomfield:2012ff}. The terms in the square brackets are the only ones that affect the dynamics of the background, and therefore the only ones of interest for our analysis. As such, we have not written explicitly all the quadratic and higher order operators that describe the dynamics of perturbations, rather collecting them into  $S^{(2)}_{\rm DE}$. Finally, $S_{m}$ is the action for all matter fields, that in the Jordan frame are minimally coupled to the metric. For a detailed explanation of how action~(\ref{eq:theaction}) is constructed we refer the reader to~\cite{Gubitosi:2012hu,Bloomfield:2012ff,Piazza:2013coa}. Here, let us stress that  this action encompasses all single-field dark energy and modified gravity models, including the $4D$ effective regime of higher dimensional theories. Given one of these models, it is possible to translate it into the EFT formalism by finding the appropriate matching for the EFT functions as elucidated in~\cite{Gubitosi:2012hu}-\cite{Bloomfield:2013efa}. The important difference of~(\ref{eq:theaction}) w.r.t. EFT of inflation~\cite{Cheung:2007st,Weinberg:2008hq} is in the conformal factor $\Omega$, which cannot be reabsorbed by a redefinition of the metric tensor because of the presence of matter.  

Varying the background action with respect to the metric and assuming a spatially flat FLRW metric one obtains the following equations:
\begin{eqnarray}
\label{Friedmann1}
&&3m_0^2\Omega H^2+3m_0^2H \dot{\Omega}=\sum_i\rho_i-\Lambda+2c,\\
\label{Friedmann2}
&&3m_0^2H^2\Omega+2m_0^2\dot{H}\Omega+m_0^2\ddot{\Omega}+2m_0^2H\dot{\Omega}=-\sum_ip_i-\Lambda,
\end{eqnarray}
where the dot indicates derivation with respect to time and $\rho_i$ and $p_i$ are, respectively, the background energy density and pressure of the $i^{th}$ matter component, for which we assume a perfect fluid form. We will consider two distinct components, i.e. dust with zero pressure (that we will indicate with a subscript `m') and radiation with $p_r=1/3\,\rho_r$.
Their continuity equations read:
\begin{eqnarray}\label{m_cons}
\dot{\rho}_m&=&-3H\rho_m,\label{conseqm}\\
\label{r_cons}
\dot{\rho}_r&=&-4H\rho_r.
\end{eqnarray}
Deriving Eq.~(\ref{Friedmann1}) w.r.t. time and combining it with Eqs.~(\ref{Friedmann2})-(\ref{r_cons}), one obtains what can be interpreted as a continuity equation for the effective DE component 
 \begin{eqnarray}\label{DE_cons}
2\dot{c}-\dot{\Lambda}&=&3m_0^2\dot{H}\dot{\Omega}-6Hc+6m_0^2H^2\dot{\Omega}\,.
\end{eqnarray}
Equations~(\ref{Friedmann1})-(\ref{DE_cons}) are all the equations we have at our disposal to study the dynamics of the background. 

The covariant, background-independent approach that we adopt~\cite{Gubitosi:2012hu,Bloomfield:2012ff}, aims at offering a general framework to study the evolution of cosmological perturbations in a model independent way. In the latter context, it is common to fix the background history to the one of $\Lambda$CDM, or something close to that, and to focus on the dynamics of perturbations. This is justified by the fact that the cosmological concordance model is in very good agreement with current observables constraining the expansion history and that most alternative models are highly degenerate with it at the level of background dynamics, while predicting modifications at the level of perturbations. In the EFT framework this practice would translate into assuming that the background is given {\it a priori}, i.e. typically  it is  chosen to be close to the $\Lambda$CDM one, and one focuses on the coefficients of the higher order operators contained in $S^{(2)}_{\rm DE}$. If we were to fix the expansion history, we could use Eqs.~(\ref{Friedmann1}) and~(\ref{Friedmann2}) to eliminate two of the three EFT functions, typically $\Lambda(t)$ and $c(t)$. This however would still leave us with one completely undetermined function of time, $\Omega(t)$ for which we should make some arbitrary choice. In our analysis, we do not fix the expansion history, but rather we keep all the three functions free and, via a dynamical analysis of the background, we identify viable forms, as well as compatibility conditions for their time dependence in order for the model to produce an expansion history that is viable. The aim of our analysis is, given the high degree of freedom, to identify general rules of cosmological viability and compatibility for $\Omega, \Lambda, c$, that can guide us in later fixing them to some form or ans$\ddot{\text{a}}$tze when performing forecasts for large scale structure data. 

\section{Dynamical system and cosmological viability}
\label{Sec3}
In this Section we will set up the necessary ingredients to perform a dynamical analysis of the effective field theory of dark energy. We need to rewrite the equations for the background into an autonomous system of first ODEs, for which we can then study the stability around equilibrium points.  To this purpose, we introduce  the following dimensionless variables:
\be\label{variables}
x=\frac{c}{3m_0^2H^2\Omega},\,\,\,\,\,\,
y= \frac{c-\Lambda}{3m_0^2H^2\Omega},\,\,\,\,\,\,
u=\frac{\rho_r}{3m_0^2H^2\Omega},\nonumber
\ee
\be
\alpha_0 =-\frac{\dot{\Omega}}{H \, \Omega}, \dots
,\alpha_n=-\frac{\Omega^{(n+1)}}{H\Omega^{(n)}},\dots
,\lambda_0 =- \frac{\dot{c}-\dot{\Lambda}}{H(c-\Lambda)}, \dots
,\lambda_m =- \frac{(c-\Lambda)^{(m+1)}}{H\,(c-\Lambda)^{(m)}},\dots
\ee
where the indices $n,m$ are unbounded from above. Using Eqs.~(\ref{Friedmann1})-(\ref{DE_cons}), we can write the following set of first order ODEs:
\begin{subequations}
\label{syst}
\begin{eqnarray}
\frac{dx}{d\ln{a}} &=& \lambda_0 y-6x-2\alpha_0 + x\alpha_0 -(\alpha_0 +2x)\f{\dot{H}}{H^2},\label{systx}\\ 
\frac{dy}{d\ln{a}} &=& \l(\alpha_0 - \lambda_0 -2\f{\dot{H}}{H^2} \r)y,\label{systy}\\ 
\frac{du}{d\ln{a}} &=&  \l(\alpha_0 -4 -2\f{\dot{H}}{H^2} \r)u,\label{systu}\\
\frac{d\alpha_{n-1}}{d\ln{a}} &=&\l(-\alpha_{n}+\alpha_{n-1}-\f{\dot{H}}{H^2}\r) \alpha_{n-1}, \hspace{0.5cm} (n\geq 1)\label{systalpha} \\
\frac{d\lambda_{m-1}}{d\ln{a}} &=&\l(-\lambda_{m}+\lambda_{m-1}-\f{\dot{H}}{H^2}\r) \lambda_{m-1},  \hspace{0.5cm} (m\geq 1) \label{systlambda}
\end{eqnarray}
\end{subequations}
where 
\begin{equation}\label{Hdot}
\frac{\dot{H}}{H^2}=-\frac{3}{2}-\frac{3}{2}x+\frac{3}{2}y+ \alpha_0 - \frac{1}{2}\alpha_1 \alpha_0 -\frac{1}{2}u.
\end{equation}
This is a nonlinear, non-autonomous system that, however, displays a hierarchical structure in the equations for the $\alpha's$ and $\lambda's$. We will shortly describe our strategy to approach it. 

Eq.~(\ref{Friedmann1}) can be read as a constraint equation
\begin{equation}\label{constraint}
\Omega_m=\frac{\rho_m}{3{m{_0}}^2\Omega H^2}=1-x-y-u-\alpha_0,
\end{equation}
with $\Omega_m\geq0$. 
When describing the cosmology of the different points, we will consider also the following parameters:
\begin{equation}
\Omega_{\rm{DE}}=x+y+\alpha_0, \hspace{1cm} \Omega_{\rm{r}}=u, \hspace{1cm}w_{\rm{eff}}\equiv-1-\frac{2}{3}\frac{\dot{H}}{H^2}=x-\frac{2}{3}\alpha_0+\frac{1}{3}\alpha_1 \alpha_0 -y+\frac{1}{3}u,
\end{equation}
respectively the DE and radiation fractional energy density and the effective equation of state. Note that what we define the fractional density parameters, are the standard ones rescaled by the function $\Omega(t)$, as it is common to do in presence of a conformal coupling~\cite{Amendola:2006we}. 

In order to solve system~(\ref{syst}) we first need to make it autonomous. The simplest option corresponds to setting $\alpha_0,\lambda_0$ to constant and evolve only the three core equations~(\ref{systx})-(\ref{systu}); we refer to this case as the zero-th order one and analyze it sampling the space $(\alpha_0,\lambda_0)$ to find viable cosmologies. As we discuss in detail in Sec.~\ref{Seczero}, this case corresponds to assuming that $\Omega$ and $c-\Lambda$ are power laws in the scale factor.
To go beyond this zero-th order analysis, we can start exploring the hierarchy of equations~(\ref{systalpha}) and~(\ref{systlambda}), by setting $\alpha_N$ and $\lambda_M$ constant for given $N,M\geq1$. We are then left with a $(3+N+M)$-dimensional system formed by the three core equations for $\{x,y,u\}$, plus $N$ equations for $\alpha_0,\dots,\alpha_{N-1}$ and $M$ equations for $\lambda_0,\dots,\lambda_{N-1}$. We perform the dynamical analysis of this system sampling the space $(\alpha_N,\lambda_M)$ and determining the regions for which one can obtain viable expansion histories.  What is the corresponding form of the EFT functions that we explore at this order? Let us develop the following argument in terms of $\Omega$; it is then trivial to reproduce it for $c-\Lambda$. 
From the definition of the $\alpha's$, we see that fixing $\alpha_N={\rm const}$ gives
\ba\label{nth_der}
\Omega^{(N)}(t)=\Omega^{(N)}(t_0)a^{-\alpha_N},
\ea
where $t_0$ is the present time. 
Now that we have an expression for the $N^{th}$ derivative of $\Omega$, we can use it to write
\ba\label{Taylor_exp}
\Omega(t)=\sum_{i=0}^{N-1}\frac{\Omega^{(i)}(t_0)}{i!}(t-t_0)^i + \Omega^{(N)}(t_0)\int_{t_0}^t\frac{(t-\tau)^{N-1}}{(N-1)!}a^{-\alpha_N}(\tau)\,d\tau,
\ea
that shows that the constant $\alpha_N$ ($N \geq 1$) parametrizes the remainder in a Taylor expansion of order $N-1$ around the present time for the function $\Omega(t)$. Notice that in order for the above argument to hold one does not necessarily need $t_0$ to be the present time (with $a_0=1$); the latter can be the desired choice in view of constraining the form of the EFT functions at recent times~\cite{Gubitosi:2012hu}, where they are expected to have a non-trivial dynamics and where they are more likely to be probed. However, one can in principle choose any other $t_0$ that is suited to one's purpose, as long as $a$ is rescaled by $a_0$ in~(\ref{nth_der}) and~(\ref{Taylor_exp}). In the following Sections, we separately analyze the stability of the system~(\ref{syst}) at different orders. In particular, after analyzing the zero-th order case in Sec.~\ref{Seczero}, we maintain $\lambda_0$ constant and focus on the $\alpha$ channel of the system, solving $3+N$ equations for the variables $\{x,y,u,\alpha_0,\dots,\alpha_N\}$. In other words, we focus on the class of models for which $c-\Lambda$ is a power law in the scale factor, while the conformal factor $\Omega$ can be increasingly general as we go up with the order. 
Alternatively one could fix $\Omega$ to a constant and open the $\lambda$ channel, which would correspond to exploring all minimally coupled models of DE. Finally, one could work with both channels and, for instance, explore, within this parametrized framework the full class of Horndeski theories~\cite{Horndeski:1974wa}. While we leave the former, as well as the most general case, for future work, we want to stress that the machinery  set up in this paper is quite general and easily applicable to the other cases mentioned above. 

Finally, let us point out that the structure of the system is such that the planes $y=0$, $u=0$, $\alpha_i=0$, $\lambda_j=0$ are all invariant manifolds, which implies that trajectories starting on one of these planes remain on it. This ensures that viable trajectories identified at a given order, will exist at all higher orders. We exploit this feature at the end of this Section, when we reconstruct the dynamics at a generic order $N\geq3$. 

\subsection{Stability Analysis}\label{Secstability}
The dynamics of system~(\ref{syst}) can be studied analyzing the evolution around fixed/critical points, i.e.  points $p_i$ satisfying the equilibrium condition $dp_i/d\ln{a}=0$. In the following we briefly summarize the general procedure; for an exhaustive description of the technique, and for some applications to cosmological models we refer the reader 
to~\cite{bookcosmol}-\cite{Baccigalupi:2000je}.
After determining its fixed points, one proceeds to calculate the eigenvalues $\mu_i$ of the Jacobian matrix $\mathcal{M}$ of the system in order to linearize it around each critical point. 
This determines the stability nature of the point, in other words it controls how the system behaves when approaching the point. We are interested in hyperbolic critical points, since around these the linearized dynamical system is a good approximation of the full nonlinear system. By definition a  critical point is said to be hyperbolic if all eigenvalues of $\mathcal{M}$ have $Re(\mu_i)\neq0$. Hyperbolic critical points are robust, in the sense that small perturbations do not change qualitatively the phase portrait near the equilibrium.  For an n-dimensional system one has n eigenvalues for each point and the stability depends on the nature of these eigenvalues, according to the following classification:
\begin{itemize}[leftmargin=*]
\item All $\mu_i$ are real and have the same sign:
\begin{itemize}
\item Negative eigenvalues $\rightarrow$ Stable node/ Attractor;
\item Positive eigenvalues $\rightarrow$ Unstable node;
\end{itemize}
\item All $\mu_i$ are real and at least one positive and one negative $\rightarrow$  Saddle points;
\item At least one eigenvalue is real and there are pairs of complex eigenvalues: 
\begin{itemize}
\item All eigenvalues have negative real parts $\rightarrow$ Stable Focus-Node;
\item All eigenvalues have positive real parts $\rightarrow$ Unstable Focus-Node;
\item At least one positive real part and one negative $\rightarrow$ Saddle Focus. 
\end{itemize}
\end{itemize} 
A working cosmological model needs to first undergo a radiation dominated era, followed by a matter era, and then enter a phase of accelerated expansion (DE) as indicated by observations~\cite{Riess:1998cb,Perlmutter:1998np}. In terms of critical points we need two saddle points for the radiation and the matter dominated eras, followed by a late time accelerated attractor, i.e. a stable node with $w_{\rm eff}<-\f{1}{3}$. 
In addition we impose the constraints that $\Omega_m\geq 0$ and $\Omega_{r}\geq 0$, given that  matter and radiation energy densities should be positively defined, and $\Omega(t)>0$ to guarantee a stable gravity~\cite{Nariai,Gurovich:1979xg}. On the other hand, we allow the effective dark energy density to be negative since this quantity may not correspond to the energy density of an actual fluid, and may indeed be negative in some models of modified gravity \cite{Amendola:2006we}.
Finally, in reconstructing viable trajectories, we require that the matter era is long enough to allow for structure formation.
\subsection{${\rm Zero}^{\rm th}$ order analysis}
\label{Seczero}
The simplest option to make the system~(\ref{syst}) autonomous is setting $\alpha_0$ and $\lambda_0$ to constant and evolve only the core equations~(\ref{systx})-(\ref{systu}). The corresponding behavior of the EFT functions is
\begin{align}
\Omega(t)=\Omega_0\, a^{-\alpha_0},\hspace{1cm} c(t)-\Lambda(t)=(c-\Lambda)_0 a^{-\lambda_0},
\end{align}
where the constants will depend on the initial conditions and their value does not affect the stability analysis. 

Unless $\alpha_0=0$, the system~(\ref{systx})-(\ref{systu}) is not closed due to the dependence on $\alpha_1$ through $\dot{H}/H^2$. We can use~(\ref{systalpha}) for $n=1$ to get 
\be
\frac{\dot{H}}{H^2}=\frac{1}{2-\alpha_0}\left(2\alpha_0 - \alpha_0^2 - 3 + 3y - 3x -u \right)\,.
\ee
The resulting critical points of the system and the analysis of their stability are shown in Table~\ref{zeroordercp}. We find that the same results are still valid if $\alpha_0=0$. In what follows we present their eigenvalues and discuss the cosmological viability.
\begin{itemize}[leftmargin=*]
\item 	$P_1$:  \underline{\textsl{matter point} } \\
The eigenvalues and the relative eigenvectors of the linearized system  around the first critical point are:
		\begin{align}
		\label{Eq:ZeroOrderP1}
		& \mu_1=-1, & &\mu_2= \alpha_0 -3, & & \mu_3=3-\lambda_0. \nonumber \\
		& \vec{u}_1=\left(\frac{\alpha_0 }{6-3 \alpha_0 }\,,\,0\,,\,1\right), & & \vec{u}_2=\left( 1\,,\,0\,,\,0\right),				& & \vec{u}_3= \left(\frac{\alpha_0 -\lambda_0 }{\alpha_0 +\lambda_0 -6}\,,\,1\,,\,0\right).
		\end{align}		
This point displays a scaling solution for which matter and DE coexists with a constant ratio between their energy densities. We are primarily interested in the matter configuration, since this is the only critical point of the zero-th order system that can provide a matter dominated critical point. If we require $\Omega_m\approx1$, then we have $\alpha_0\approx0$, which combined with the requirements of having a saddle, gives $\alpha_0=0 \wedge \lambda_0 < 3$. 
\item 	$P_2$:  \underline{\textsl{stiff matter point} }
		\begin{align}
		\label{Eq:ZeroOrderP2}
		& \mu_1=2-\alpha_0 , & & \mu_2= 3-\alpha_0, & & \mu_3=-\alpha_0 -\lambda_0 +6. \nonumber \\
		& \vec{u}_1=(-1\,,\,0\,,\,1), & & \vec{u}_2=(1\,,\,0\,,\,0), & & \vec{u}_3=(-1\,,\,1\,,\,0).
		\end{align}
This point is a DE dominated critical point; it is a stable node with accelerated expansion only if $\alpha_0 >3\wedge	\alpha_0 +\lambda_0 >6$. 
For $\alpha_0=0$, it has $w_{\rm eff}=1$, which corresponds to a stiff matter behavior that could be of interest for modeling early stages of the Universe~\cite{Copeland:1997et}.
\item 	$P_3$:  \underline{\textsl{DE point}}
		\begin{align}
		\label{Eq:ZeroOrderP3}
		& \mu_1=\lambda -4, & & \mu_2= \lambda_0 -3, & & \mu_3=\alpha_0 +\lambda_0 -6. \nonumber\\
		& \vec{u}_1=\left(\frac{\lambda_0 -2 \alpha_0 }{3 (\alpha_0 -2)}\,,\,-\frac{\alpha_0 +\lambda_0 -6}{3 (\alpha_0 -2)}\,,\,1\right) ,& & \vec{u}_2= \left(\frac{\alpha_0 -\lambda_0 }{\alpha_0 +\lambda_0 -6}\,,\,1\,,\,0\right),& &\vec{u}_3=\left(-1\,,\,1\,,\,0\right).
		\end{align}
This is the second DE dominated critical point of the zero-th order system; it exhibits a correct cosmological behavior, i.e. $w_{\rm eff}<-\f{1}{3}$, if $(\alpha_0 \geq 3\land \alpha_0 +\lambda_0 <6)\lor(\alpha_0 <1\land \lambda_0 <\alpha_0 +2)\lor (1\leq\alpha_0 <3\wedge \lambda_0 <3)$.
\item 	$P_4$:  \underline{\textsl{radiation point} }
		\begin{align}
		\label{Eq:ZeroOrderP4}
		& \mu_1=1, & & \mu_2= \alpha_0 -2, & & \mu_3=4-\lambda_0. \nonumber \\
		& \vec{u}_1=\left( \frac{\alpha_0 }{6-3 \alpha_0 }\,,\,0\,,\,1\right), & & \vec{u}_2=\left(-1\,,\,0\,,\,1\right), & &\vec{u}_3=					\left(\frac{\lambda_0 -2 \alpha_0 }{3 (\alpha_0 -2)}\,,\,-\frac{\alpha_0 +\lambda_0 -6}{3 (\alpha_0 -2)}\,,\,1\right).
		\end{align}
This point is characterized by $\Omega_m=0$ and a coexistence of radiation and DE with a constant energy density ratio; in other words it is a scaling radiation point. We will focus on its radiation dominated version, since it is the only point that can supply a radiation era for the zero-th order trajectories. It can be  be easily seen that it corresponds to a saddle with $w_{\rm eff}=\f{1}{3}$ if $\alpha_0 =0\land \lambda_0\neq 4$.
\end{itemize}
Combining all the information above,  we conclude that viable cosmological models for the zero-th order case, can be recovered setting $\alpha_0=0$ and $\lambda_0 < 3$, and they are characterized by the transitions $P_4 \rightarrow P_1 \rightarrow P_3 $ (radiation$\rightarrow$ matter $\rightarrow$ DE attractor). One can actually further constrain the space 
$(\alpha_0,\lambda_0)$. Indeed, a peculiar feature of the zero-th order system is the disposition of the critical points.
A careful analysis of the eigenvectors~(\ref{Eq:ZeroOrderP1})-(\ref{Eq:ZeroOrderP4}), shows that for any pair of critical points  the heteroclinic orbits, i.e. the lines connecting the two points, are strainght lines. This is valid for any choice of $(\alpha_0,\lambda_0)$ and it allows us to put a stricter bound on $\lambda_0$ by requiring a long enough matter era for the trajectories of interest. Let us elucidate this point.
The $\Lambda$CDM model corresponds to $\alpha_0=\lambda_0=0$ and its trajectory is such that it starts very close to the radiation saddle point $P_4$, then it passes close to the matter  saddle $P_1$ and finally it reaches the dark energy attractor $P_3$, always moving very close to the heteroclinic orbits that connect these three critical points. The time spent by this trajectory in the last transition gives a handle on the proper duration of the matter era for trajectories that aim to be cosmologically viable. Since after we set $\alpha_0=0$ the coordinates of $P_1$ and $P_4$ are fixed (i.e. independent on $\lambda_0$), we can use the constraint on the position of $P_3$ to put a stringent bound on $\lambda_0$; indeed if we change the latter, and hence move $P_3$,  the duration of the matter era will change significantly. In other words, we need $P_3$ to be always close to its $\Lambda$CDM position, and this forces $\lambda_0\sim 0$. \\
In summary, \emph{viable cosmological models for the zero-th order case, can be recovered setting} $\alpha_0=0$ and $\lambda_0 \approx 0$, \emph{and they are characterized by the transitions} $P_4 \rightarrow P_1 \rightarrow P_3 $. Notice that $\alpha_0=0$ implies that the conformal factor $\Omega(t)$ is a constant, which just rescales the Planck mass.
\begin{table}
\centering
\footnotesize
\renewcommand\arraystretch{1.7}
\begin{tabular}{cp{5cm}p{6cm}p{2.1cm}p{2.1cm}}
\hline
\hline
\centering{Point}&\centering{$[x_c,y_c,u_c]$} & \centering{Stability} & \centering{$\Omega_{\rm DE}$} & $w_{\rm{eff}}$  \\
\hline
\hline
\centering{$P_{1}$}&\centering{$\l[-\frac{1}{6}\alpha_0(1+\alpha_0),0,0\r]$} & \centering{{\bf Stable node:} $\lambda_0 > 3 \wedge \alpha_0 \le 2$ \\{\bf Saddle point:} $(\lambda_0 < 3 \wedge \alpha_0\le 2)\vee(\alpha_0>3\wedge \lambda\neq 3)$} &\centering{$-\frac{1}{6} (\alpha_0 -5) \alpha_0$} & $-\frac{\alpha_0 }{3}$      \\
\hline 
\centering{$P_{2}$}&\centering{$\l[1-\alpha_0,0,0\r]$}  & \centering{{\bf Stable node:} $\alpha_0>3\wedge\alpha_0 + \lambda_0>6$. \\ {\bf Unstable node:} $\alpha_0<2\wedge\alpha_0 + \lambda_0 <6$ \\ {\bf Saddle point:} otherwise}  &\centering{1} &  $1-\frac{2 \alpha_0 }{3}$	   \\
\hline
\centering{$P_{3}$}&\centering{$\l[\frac{1}{12} (-\alpha_0^2-\alpha_0(\lambda_0 +4)+2 \lambda_0)\r.$,\\ $\l.\frac{1}{12} (\alpha_0 -2) (\alpha_0 +\lambda_0 -6),0\r]$} & \centering{{\bf Stable node:} $(\alpha_0 \geq 3 \wedge \alpha_0 +\lambda_0 <6)\vee (\alpha_0 <3\wedge \lambda_0 <3)$.\\ {\bf Unstable node:} $(\lambda_0 >4\wedge \alpha_0 \geq 2)\vee (\alpha_0 +\lambda_0 >6\wedge \alpha <2)$.\\ {\bf Saddle point:} otherwise} &\centering{1}  & $\frac{1}{3} (-\alpha_0 +\lambda_0 -3)$  \\
\hline 
\centering{$P_{4}$}&\centering{$\l[-\frac{\alpha_0 ^2}{4},0,\frac{1}{4} (\alpha_0 -2)^2\r]$} & \centering{{\bf Unstable node:} $\alpha_0 >2\wedge \lambda_0 <4$ \\ {\bf Saddle point:} $(\alpha_0 >2\wedge \lambda_0 >4)\vee (\alpha_0 <2\wedge \lambda_0 \neq 4)$} & \centering{$-\frac{1}{4} (\alpha_0 -4) \alpha_0$ }& $\frac{1-\alpha_0 }{3}$ \\
\hline
\hline
\end{tabular}
\renewcommand\arraystretch{1}
\caption{Hyperbolic critical points and stability analysis for the \emph{zero-th order} system. The additional constraints $\Omega_m\geq0$ and $\Omega_r\geq 0$ have been imposed.  We have $\mathcal{D}\equiv\left\{  \alpha_0 \mbox{,}  \lambda_0  \in \mathbb{R} \right \}$.}\label{zeroordercp}
\end{table}
\subsubsection{Reconstructing quintessence models}\label{quintessence}
\begin{figure}[tbp!]
\centering
\includegraphics[width=17.5cm]{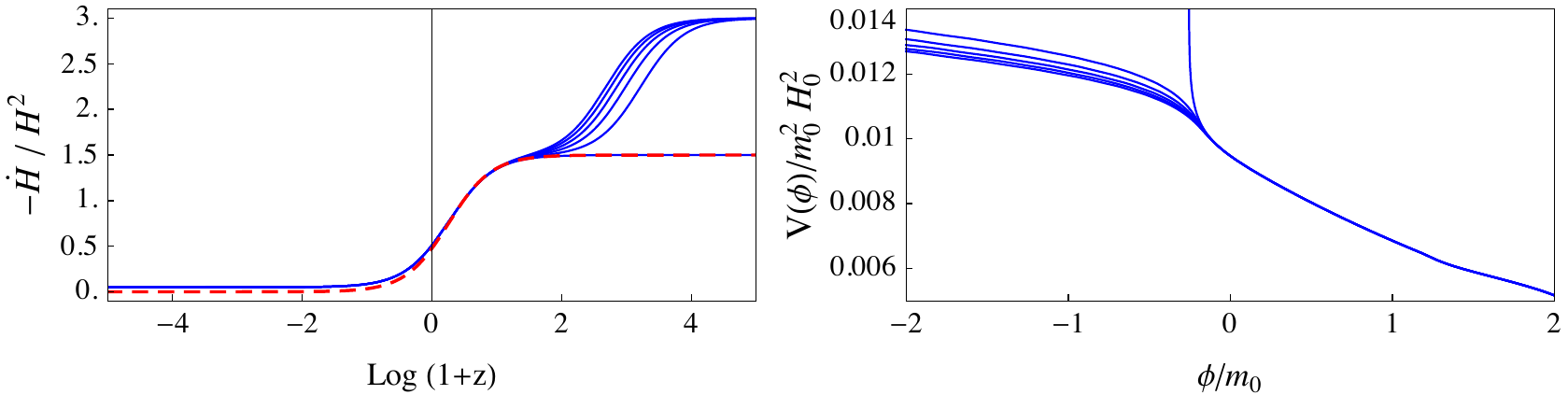}
\caption{The slow roll parameter and the quintessence potential reconstructed for several trajectories of the $\alpha_0=0$, $\lambda_0=0.1$ model (blue lines). The red dashed line represents the behavior of the Planck~\cite{Ade:2013zuv} best fit $\Lambda$CDM model.}
\label{Fig:ZeroOrderQuint}
\end{figure}
We shall now show how the results of this general dynamical analysis can be reverberated to constrain specific models of dark energy. As an example, we choose to interpret the results of the zero-th order analysis within the context of quintessence by using the matching in~\cite{Gubitosi:2012hu,Bloomfield:2012ff}. Given that $c$ and $\Lambda$ for quintessence models assume the following forms:
\begin{align}
\label{Eq:QuintMatch}
c= \frac{\dot{\phi}^2}{2}\,, \hspace{1cm} c-\Lambda = V(\phi)=(c-\Lambda)_0 a^{-\lambda_0},
\end{align}
one immediately notices that the bound $\alpha_0=0$ for the zero-th order analysis, translates into the constraint that any quintessence model with a potential which is a power law in the scale factor, cannot have a power law behavior for the conformal factor $\Omega$, and therefore at this order is forced to be minimally coupled.
Among the models selected in this way, we will choose for our example those corresponding to the value $\lambda_0=0.1$. For simplicity we do not include radiation in this numerical study since it will not alter much the reconstruction. We  choose initial conditions so that the present day matter density matches the Planck  $\Lambda$CDM best fit value~\cite{Ade:2013zuv} and we sample the trajectories that undergo a $ P_1 \rightarrow P_3$ transition. 
Then using~(\ref{Eq:QuintMatch}) we reconstruct the time evolution of the quantities of interest, i.e. the slow-roll parameter and the the potential. We show the outcome in Fig.~\ref{Fig:ZeroOrderQuint}, where one can notice that the late time DE attractor corresponds to slow roll behavior which makes the field behave approximately like a cosmological constant. On the other hand at early times the field is rolling down the potential very fast as the dark energy component behaves like stiff matter, as expected since the unstable stiff-matter point, $P_2$, serves as the starting point for the numerically reconstructed trajectories. The corresponding potential is monotonically decreasing and positively defined. 
\subsection{First order analysis}
\label{Secfirst}
\begin{figure}[t]
\centering
\includegraphics[width=17.5cm]{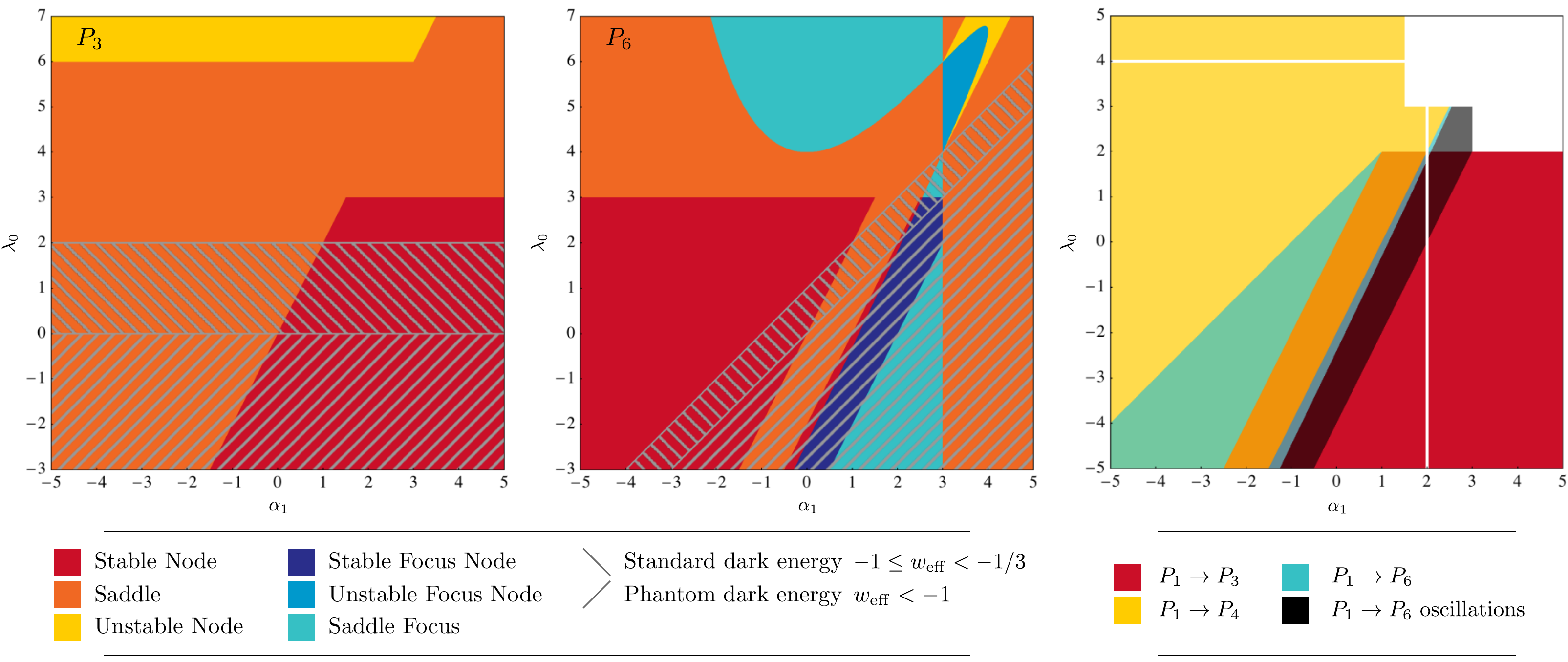}
\caption{The left panel shows the results of the stability analysis of the \emph{first order} system around $P_3$. The panel at the center illustrates the stability  around $P_6$.
The right panel shows the combined results of the \emph{first order} analysis: regions in the $(\alpha_1 , \lambda_0)$ plane which allow the different transitions discussed in Sec.~\ref{Sub:FirstOrderNumerical} are shown in different colors.}
\label{Fig:FirstOrderRegions}
\end{figure}

We now start exploring the hierarchy of equations for the $\alpha's$. The immediate generalization of the previous model is the one obtained by letting $\alpha_0$ vary, while fixing ($\alpha_1, \lambda_0$) to constant. As discussed at the beginning of this Section, this corresponds to setting 
\begin{align}
\dot{\Omega}(t)=\dot{\Omega}_0a^{-\alpha_1}, & &c(t)-\Lambda(t)=(c-\Lambda)_0 a^{-\lambda_0},
\end{align}
where again the constants will depend on the initial conditions and do not affect the stability analysis.  
Our system of equations is now formed by~(\ref{systx})-(\ref{systu}) along with Eq.~(\ref{systlambda}) with $n=1$ and the constraint~(\ref{constraint}). The system has nonlinear quadratic terms and, for different values of  the parameters ($\alpha_1$, $\lambda_0$), it can display a wide range of behaviors. 
\begin{table}
\centering
\footnotesize
\renewcommand\arraystretch{1.7}
\begin{tabular}{cp{5cm}p{5cm}p{2.8cm}{c}}
\hline
\hline
\centering{Point} &\centering{$\l[x_c,y_c,u_c,\alpha_{0,c}\r]$}& \centering{Stability} & \centering{$\Omega_{\rm DE}$} & $w_{\rm{eff}}$  \\
\hline
\hline
\centering{$P_{1}$}&\centering{$\l[0,0,0,0\r]$} & \centering{{\bf Stable node:} $\lambda_0 > 3\wedge \alpha_1>\frac{3}{2} $ \\ {\bf Saddle point:} otherwise}   & \centering{0} &  0         \\
\hline
\centering{$P_{2}$}&\centering{$\l[1,0,0,0\r]$}& \centering{ {\bf Unstable node:} $\alpha_1<3 \wedge \lambda_0<6 $ \\ {\bf Saddle point:} otherwise} & \centering{1} & 1  \\
\hline
\centering{$P_{3}$}&\centering{$\l[\frac{\lambda_0}{6}, 1-\frac{\lambda_0}{6}, 0,0\r]$}   & \centering{{\bf See Fig.~\ref{Fig:FirstOrderRegions}. }} &\centering{1} & $\frac{1}{3}(\lambda-3)$  \\
\hline
\centering{$P_{4}$}&\centering{$\l[-1,0,0,2\r]$} &  \centering{ {\bf Stable node:} $\alpha_1<\frac{5}{2} \wedge 2-2\alpha_1+\lambda_0 > 0 $,\\ {\bf Unstable node:} $ \alpha_1>3 \wedge 2-2\alpha_1+\lambda_0<0$, \\ {\bf Saddle point:} otherwise}  & \centering{1}&  $\frac{1}{3}(-7+2\alpha_1)$ \\
\hline
\centering{$P_{5}$}& \centering{$\l[\frac{1}{3}(-3+5\alpha_1-2\alpha_1^2),0,0,-3+2\alpha_1\r]$} &  \centering{{\bf Stable node:}\hspace{5cm} $\alpha_1<\frac{3}{2} \wedge \lambda_0>3$ \hspace{5cm} {\bf Saddle point:} \hspace{5cm} $\alpha_1<\frac{3}{2} \wedge \lambda_0<3 \,\, \vee \,\, \alpha_1>3 \wedge \lambda_0>3$ \hspace{5cm} $\vee\, \alpha_1>3 \wedge  \lambda_0<3\, \vee  \hspace{5cm}\lambda_0\neq 3 \wedge \frac{3}{2}<\alpha_1<\frac{5}{2} $ }   & \centering{ $-4+\frac{11\alpha_1}{3}-\frac{2\alpha_1^2}{3}$} & $1-\frac{2\alpha_1}{3} $   \\
 \hline
\centering{$P_{6}$}&\centering{$\l[\frac{1}{6}[-2\alpha_1^2+\alpha_1(\lambda_0-4)+3\lambda_0],\r.$\\$\l. \frac{1}{6}(-3+\alpha_1)(-2+2\alpha_1-\lambda_0),\r.$\\$\l. 0,2\alpha_1-\lambda_0\r]$} &  \centering{ \bf See  Fig.~\ref{Fig:FirstOrderRegions}. }
 & \centering{ 1} &  $\frac{1}{3}(-3-2\alpha_1+2\lambda_0)$  \\
 \hline
\centering{$P_{7}$}& \centering{$\l[0,0,1,0\r]$}&  \centering{{\bf Saddle point:} $\lambda_0 \neq 4 \wedge \alpha_1\neq2  $}   & \centering{ 0 } & $\frac{1}{3} $         \\
 \hline
\centering{$P_{8}$}&\centering{$\l[-(-2+\alpha_1)^2,0,(-3+\alpha_1)^2,2(-2+\alpha_1)\r]$} &  \centering{ {\bf Saddle point:}  $\alpha_1 \neq 3  \wedge \alpha_1 \neq 2 \wedge \lambda_0\neq 4  $} &  \centering{$-8+6\alpha_1-\alpha_1^2$} & $\frac{1}{3}(5-2\alpha_1)$\\
\hline
\hline
\end{tabular}
\renewcommand\arraystretch{1}
\caption{Hyperbolic critical points of the \emph{first order} analysis ($\alpha_1,\lambda_0={\rm constant}$), for which we have imposed the additional constraints $\Omega_{m}\geq0$ and $\Omega_{r}\geq0$. We have $\mathcal{D}\equiv\left\{  \alpha_1 \mbox{,}  \lambda_0  \in \mathbb{R} \right\}$.} \label{tabfirstorder}
\end{table}
The critical points  of the system and the stability properties according to their eigenvalues are summarized  in Table~\ref{tabfirstorder}. In what follows we give a more detailed overview of each point, reporting the corresponding eigenvalues. 
\begin{itemize}[leftmargin=*]
\item $P_1$: \underline{ \textsl{matter point}} \\
The linearized system around the first critical point has the following eigenvalues: 
\begin{equation}
\mu_1=-3, \,\,\,\, \mu_2= \frac{3}{2}-\alpha_1, \,\,\,\,  \mu_3= -1, \,\,\,\, \mu_4=3-\lambda_0 .
\end{equation}
It corresponds to a matter dominated solution ($w_{\rm{eff}}=0$) which is a saddle point for $\lambda_0 \neq 3\wedge \alpha_1<\frac{3}{2} \vee \lambda_0 < 3\wedge \alpha_1 >\frac{3}{2}$.

\item $P_2$: \underline{ \textsl{stiff matter point} }
\begin{equation}
\mu_1=3, \,\,\,\, \mu_2= 3-\alpha_1, \,\,\,\, \mu_3=6-\lambda_0, \,\,\,\, \mu_4=2.
\end{equation}
This point corresponds to unstable solutions  with a stiff matter equation of state, which could be relevant in the early stages of the Universe~\cite{Copeland:1997et}. 

\item  $P_3$:  \underline{\textsl{DE point} }
\begin{equation}
\mu_1=-6+\lambda_0, \,\,\,\, \mu_2= -3+\lambda_0, \,\,\,\, \mu_3=-\alpha_1+\frac{\lambda_0}{2}, \,\,\,\, \mu_4=-4+\lambda_0.
\end{equation}
It gives a DE dominated solution which is accelerated for $\lambda_0<2$. For $\lambda_0<0$ the point has a phantom equation of state.
In particular we have a late time accelerated attractor (i.e. a stable node), with $a \propto t^{2/\lambda_0} $, for $(\alpha_1>1 \wedge \lambda_0<2)  \vee (\lambda_0<2\alpha_1 \wedge \alpha_1\leq1)$.

\item $P_4$:  \underline{\textsl{phantom DE point} }
\begin{equation}
\mu_1=-5+2\alpha_1, \,\,\,\, \mu_2= -3+\alpha_1, \,\,\,\, \mu_3=-2+2\alpha_1-\lambda_0,\,\,\,\, \mu_4=2(\alpha_1-3).
\end{equation}
It has a DE dominated solution with an accelerated expansion for  $\alpha_1<3$, (with a phantom equation of state for $\alpha_1<2$). Furthermore, the point is a saddle for $\alpha_1<\frac{5}{2} \wedge \lambda_0 >  -2+2\alpha_0$ with  $a\propto t^{\frac{1}{\alpha_1-2}}$.

\item $P_5$: \underline{ \textsl{matter scaling point} }
\begin{eqnarray}
\mu_1&=&\frac{1}{4}\left( -21+13\alpha_1-2\alpha_1^2-\sqrt{81-42\alpha_1+29\alpha_1^2-20\alpha_1^3+4\alpha_1^4}  \right) ,\nonumber\\
\mu_2&=& \frac{1}{4}\left( -21+13\alpha_1-2\alpha_1^2+\sqrt{81-42\alpha_1+29\alpha_1^2-20\alpha_1^3+4\alpha_1^4}  \right) ,\nonumber \\
\mu_3&=& 3-\lambda_0, \,\,\,\, \mu_4=-1.
\end{eqnarray}
For this critical point we have a matter scaling solution with $\Omega_m=5-\frac{11}{3}\alpha_1+\frac{2}{3}\alpha_1^2$ and $\Omega_{\rm DE}= -4+\frac{11\alpha_1}{3}-\frac{2\alpha_1^2}{3}$. The constraint on the positiveness of the matter density gives $\alpha_1\geq3 \vee \alpha_1\leq\frac{5}{2}$. In this paper we do not perform a full analysis of scaling solutions, but we rather focus on the two extrema for which either of the two components has fractional energy density equal to unity. We leave the full analysis of the scaling regime for future work. For this specific point it means that we consider only the case for which $\Omega_m=1$ and the case for which $\Omega_{\rm DE}=1$. Both points do not display the proper cosmology and therefore we do not consider $P_5$ any further.

\item $P_6$: \underline{\textsl{DE point}} 
\begin{eqnarray} \label{Eq:FirstOrderP6}
\mu_1&=&\lambda_0-3, \,\,\,\, \mu_2\,\,=\,\,\lambda_0-4,\nonumber\\
\mu_3&=&\frac{1}{4}\left(-12-2\alpha_1^2-3\lambda_0+\alpha_1(10+\lambda_0)\right.\nonumber\\
&&\left.-\sqrt{-3+\alpha_1}\sqrt{-48+4\alpha_1^3-4\alpha_1^2(\lambda_0-1)-8\lambda_0+5\lambda_0^2+\alpha_1(32-12\lambda_0+\lambda_0^2)}\right) ,\nonumber\\
\mu_4&=&\frac{1}{4}\left(-12-2\alpha_1^2-3\lambda_0+\alpha_1(10+\lambda_0)\right.\nonumber\\
&&\l.+\sqrt{-3+\alpha_1}\sqrt{-48+4\alpha_1^3-4\alpha_1^2(\lambda_0-1)-8\lambda_0+5\lambda_0^2+\alpha_1(32-12\lambda_0+\lambda_0^2)}\right) .
\end{eqnarray}
The point $P_6$ gives  a DE dominated solution, with $a(t)\propto t^{\frac{1}{\lambda_0-\alpha_1}}$,  which gives an accelerated expansion for $\lambda_0-\alpha_1<1$ (phantom  if $\alpha_1>\lambda_0$).  The results of the stability analysis around this critical point are summarized in Fig.~\ref{Fig:FirstOrderRegions}; one can identify regions in the space $(\alpha_1,\lambda_0)$ for which the point is a late time attractor, as well as regions for which it is a stable focus-node. The latter one is an asymptotically stable point and corresponds to the case in which the system undergoes oscillations prior to reaching the equilibrium.

\item $P_7$: \underline{\textsl{radiation point}}
\begin{equation}
\mu_1= -2, \,\,\,\, \mu_2=1, \,\,\,\, \mu_3= -\alpha_1+2, \,\,\,\, \mu_4=4-\lambda_0.
\end{equation}
It corresponds to a standard radiation point with $w_{\rm{eff}}=\frac{1}{3}$ and can be a saddle  for $\alpha_1\neq2$ and $\lambda_0\neq4$.

\item $P_8$: \underline{\textsl{radiation scaling point}} 
\begin{equation}
\mu_1= 1, \,\,\,\, \mu_2=2(\alpha_1-3), \,\,\,\, \mu_3= -6+5\alpha_1-\alpha_1^2, \,\,\,\, \mu_4=4-\lambda_0.
\end{equation}
This point exhibits a radiation scaling behavior since $\Omega_m=0$ while $\Omega_r$ and $\Omega_{\rm DE}$ can be both non-vanishing. However one cannot in general find values of $(\alpha_1,\lambda_0)$ that give either a proper DE or radiation dominated cosmology.
\end{itemize}

As we already discussed, a working cosmological model needs to first undergo a radiation dominated era, followed by a matter dominated era (that needs to be long enough to allow for proper structure formation) and finally it has to approach an accelerated phase. The only critical point which corresponds to a proper  radiation domination in the first order system is $P_7$, which is a saddle for $\alpha_1\neq2$ and $\lambda_0\neq4$; a good critical point for a matter  era is $P_1$, which can be a saddle with $a \propto t^{2/3}$. From this point the system can move to an accelerated expansion phase by going toward the late time attractors $P_3$, $P_4$ or $P_6$ (as well as the stable-focus version of $P_6$), depending on the values of $\alpha_1,\lambda_0$. Therefore we have three types of cosmologically viable trajectories, that can be identified by the last transition that they undergo: $P_1\rightarrow P_3$, $P_1\rightarrow P_4$ and $P_1\rightarrow P_6$ (with and without oscillations). In the next subsection we investigate numerically each of these cases.
Finally, we give a graphical representation of the regions in $(\alpha_1,\lambda_0)$ for which the different transitions can take place in Fig.~\ref{Fig:FirstOrderRegions}. 
\subsubsection{Numerical investigation of different transitions}\label{Sub:FirstOrderNumerical}
We shall now investigate numerically the structure of the phase space for some models that display the different types of possible transitions discussed above. In order to facilitate the visualization of the phase space, we neglect radiation.
\begin{figure}
\centering
\subfigure[The $\alpha_1=0.1$, $\lambda_0=0$ model.\label{Fig:Phase1}]
{\includegraphics[width=7.8cm]{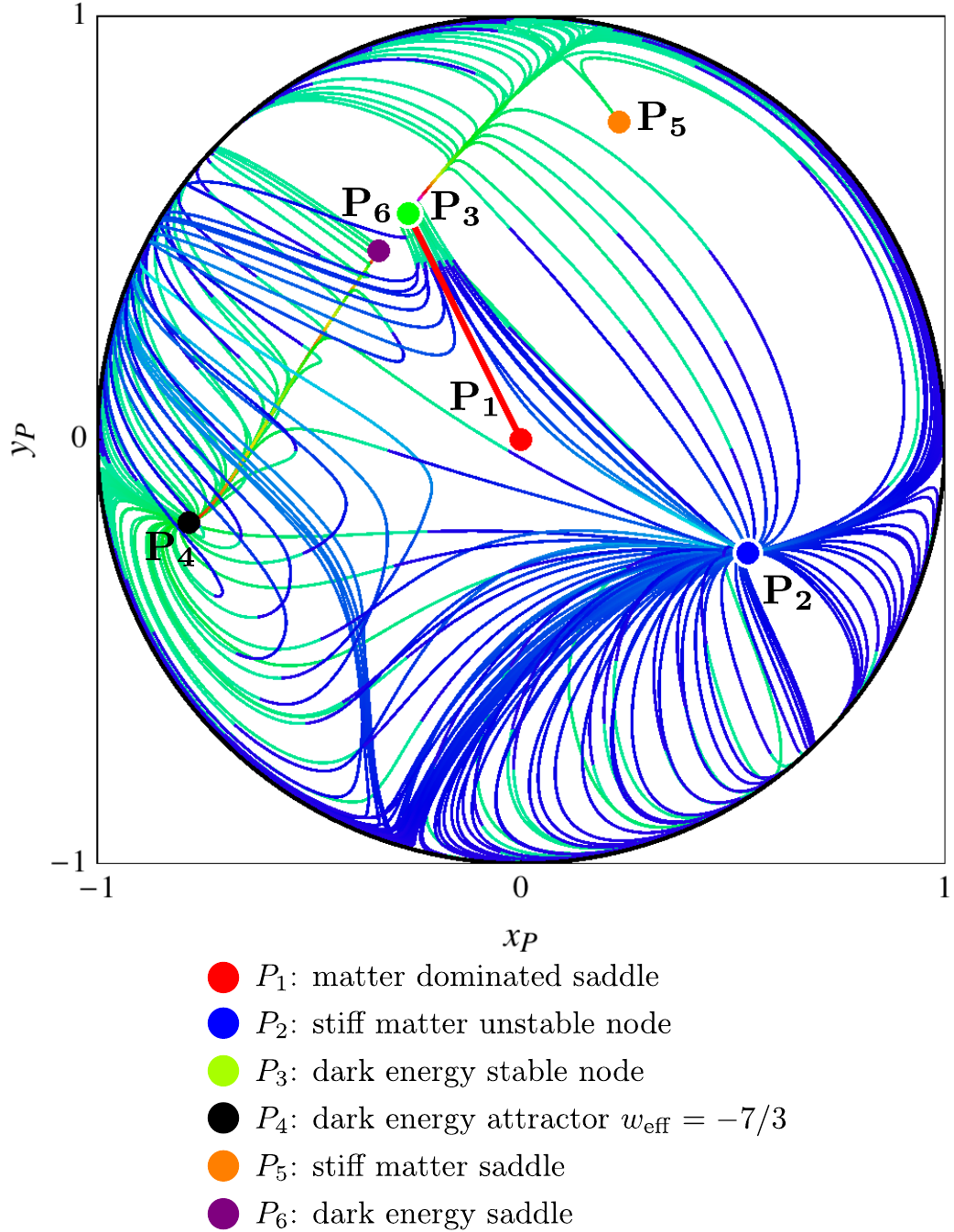}}
\hspace{2mm}
\subfigure[The $\alpha_1=2.4$, $\lambda_0=1.3$ model.\label{Fig:Phase2}]
{\includegraphics[width=7.8cm]{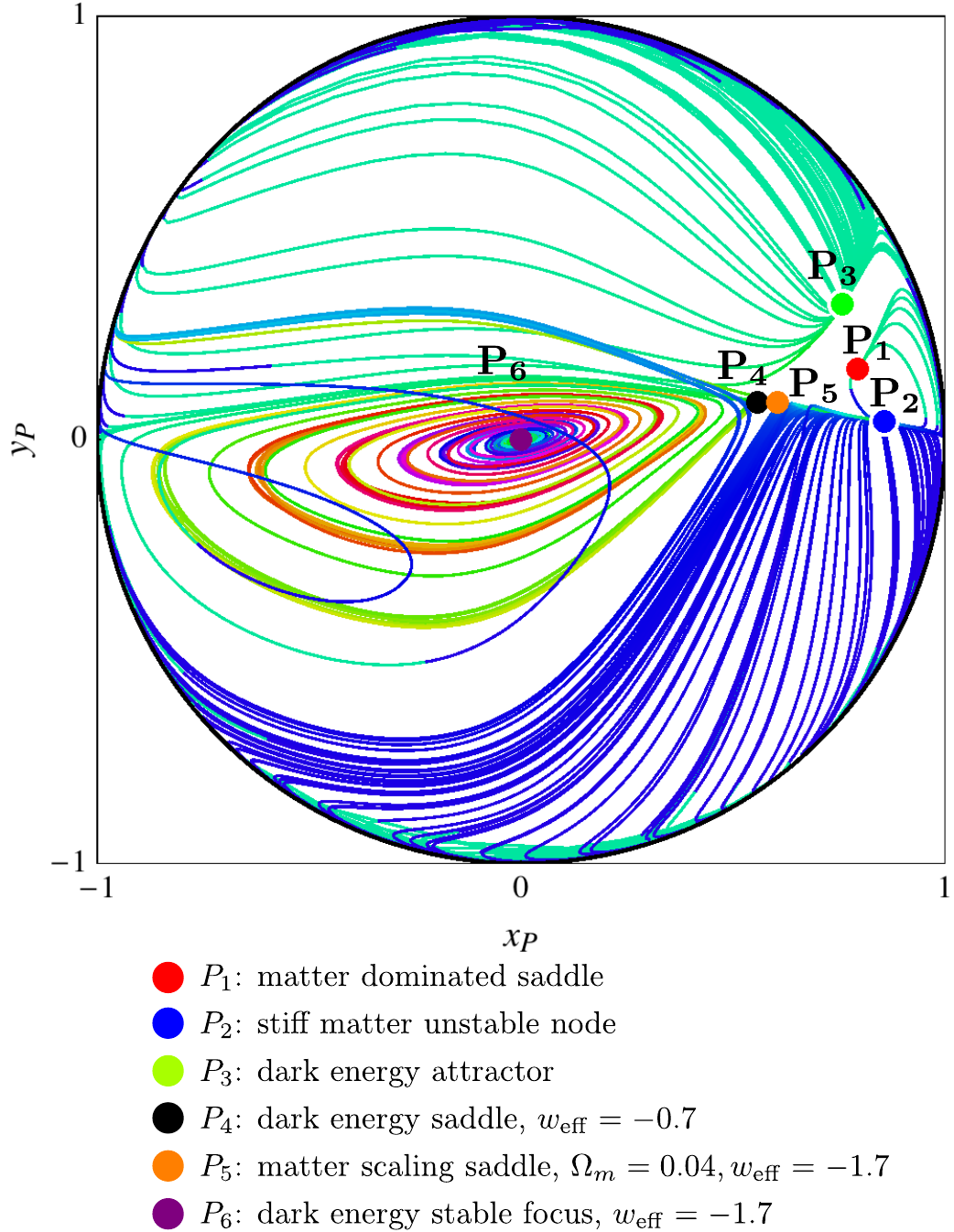}} \\
\subfigure[The $\alpha_1=0$, $\lambda_0=-1/2$ model.\label{Fig:Phase3}]
{\includegraphics[width=7.8cm]{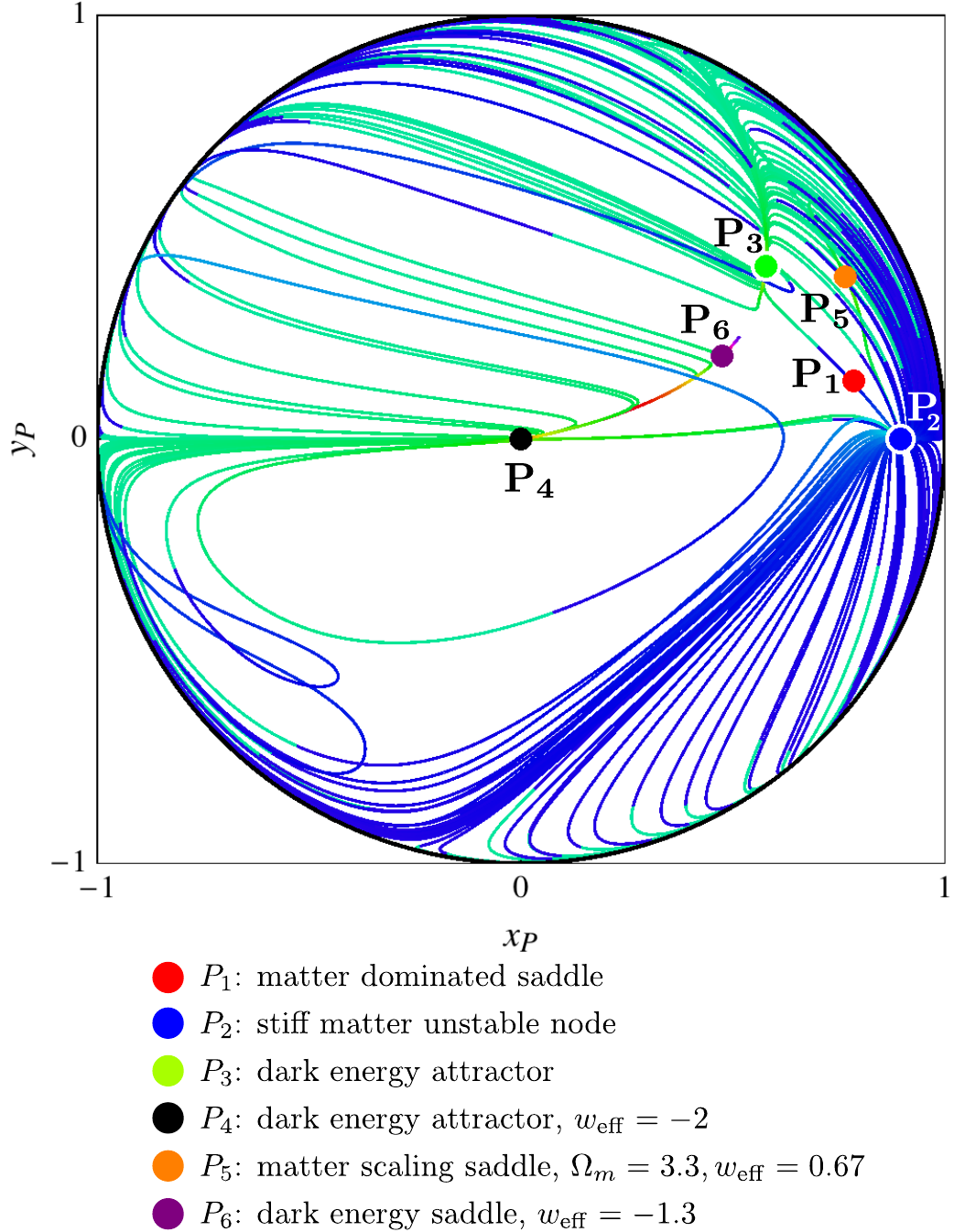}}
\hspace{2mm}
\subfigure[The $\alpha_1=-2$, $\lambda_0=-2$ model.\label{Fig:Phase4}]
{\includegraphics[width=7.8cm]{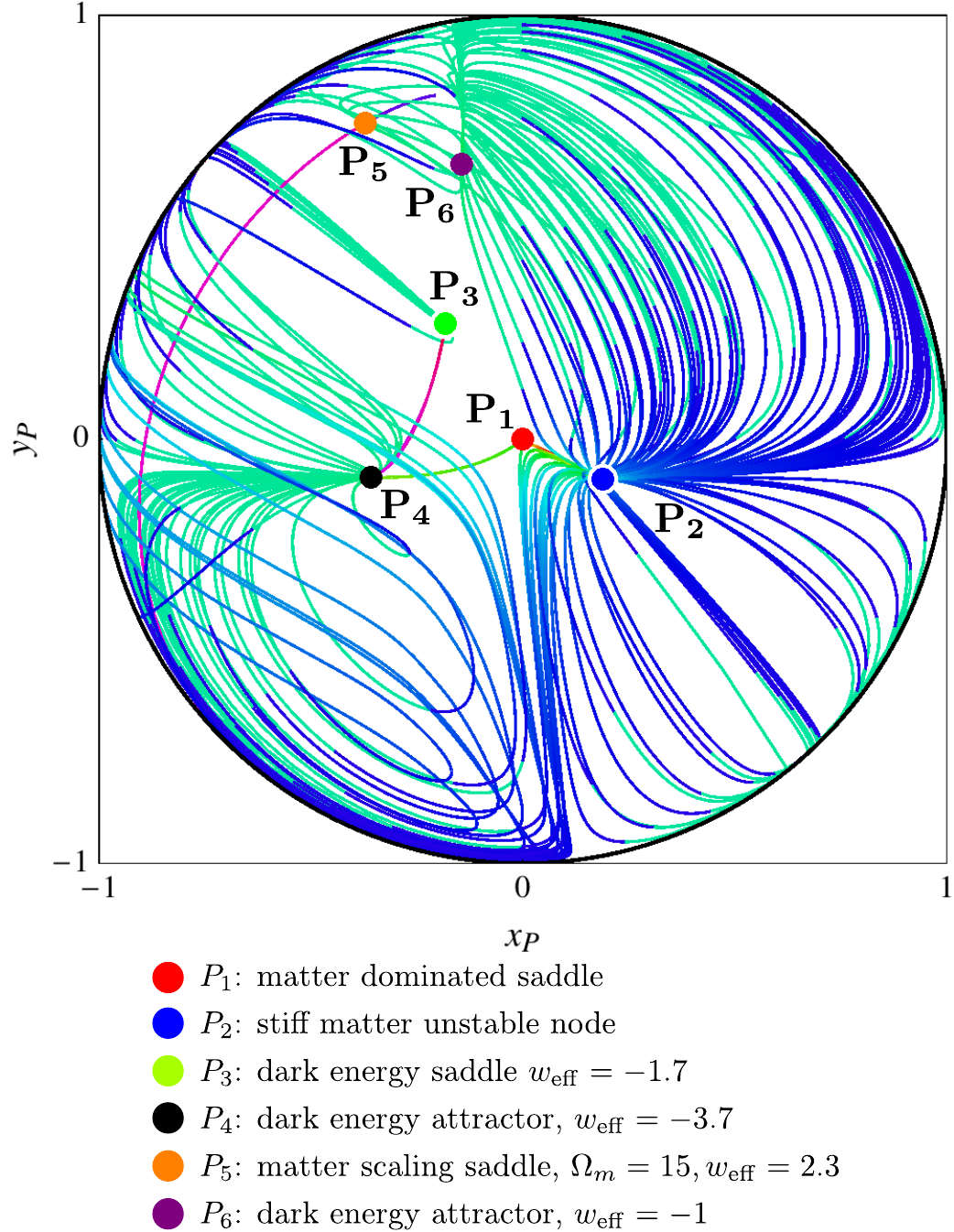}} \\
\caption{The phase space numerical investigation of different dark energy models for the \emph{first order} system. Initial conditions are evolved both in the past (blue lines) and in the future (green lines). The red line in (a) corresponds to the $\Lambda$CDM trajectory.}
\label{Fig:PhaseSpace}
\end{figure}
Let us briefly describe the procedure that we follow for this numerical investigation. 
We set initial conditions in order to reproduce the $\Lambda$CDM matter density~\cite{Ade:2013zuv} at some given initial redshift and we systematically sample trajectories that cross the plane so defined.
After the integration of the equations of motion we notice that, even if nothing a priori suggests it, the trajectories that depart from constant matter density planes remain quite close to them.
It is then possible to visualize the behavior of the three dimensional system by projecting the trajectories on these planes, and compactifying the latter via
\begin{align}
x_P=\frac{x}{\sqrt{1+ x^2 + y^2}}\,,\hspace{1cm} y_P=\frac{y}{\sqrt{1+ x^2 + y^2}}.
\end{align}
After this operation we obtain the phase space plots that are shown in Fig.~\ref{Fig:PhaseSpace}. In what follows we discuss the different types of transitions recovered with the technique just described; in particular we choose four different combinations of values for $(\alpha_1,\lambda_0)$, according to the previous analysis (e.g. Fig.~\ref{Fig:FirstOrderRegions}), to focus each time on a different type of trajectory among the cosmologically viable ones.\\

\underline{$P_1\rightarrow P_3$ transition.}
We start with the model corresponding to $\alpha_1=0.1$ and $\lambda_0=0$. This choice of values allows us to recover trajectories that mimic very closely the $\Lambda$CDM trajectory, shown as a red line in Fig.~\ref{Fig:PhaseSpace}a.
Notice that for this choice of $\alpha_1,\lambda_0$, there is an alternative stable attractor, $P_4$, which gives a phantom DE.
 We set initial conditions to reproduce $\Omega^0_{m} = 0.31$ and evolve the system to obtain the phase space plot shown in Fig.~\ref{Fig:PhaseSpace}a. One can notice that the phase space is dominated in the past by trajectories moving away from the unstable point $P_2$. These trajectories can be divided in several groups. The first one is made by trajectories that leave $P_2$ and reach infinity. Obviously these correspond to unphysical solutions since the matter density and/or $w_{\rm eff}$ would go to infinity as well. The second group is made of trajectories that leave $P_2$ to go to $P_3$ and exhibit a cosmological behavior that is very similar to the $\Lambda$CDM one.
The third family of trajectories leave $P_2$ to go to $P_4$ that is the phantom DE attractor, while the fourth family of trajectories is made up by solutions that leave infinity and go to $P_4$ and $P_3$.
It is worth noticing that we find again the $P_2 \rightarrow P_1 \rightarrow P_3$ transition that we had found for the zero-th order system. In fact, the eigenvector that corresponds to the positive eigenvalue of $P_1$ is aligned with the eigenvector that corresponds to the negative eigenvalue of $P_3$ and the same holds for $P_1$ and $P_2$. As we already discussed, this gives rise to a family of cosmologically viable trajectories (noticeable in Fig.~\ref{Fig:PhaseSpace}a) that move very close to the heteroclinic orbits connecting these points.\\
\begin{figure}[t]
\centering
\includegraphics[width=18cm]{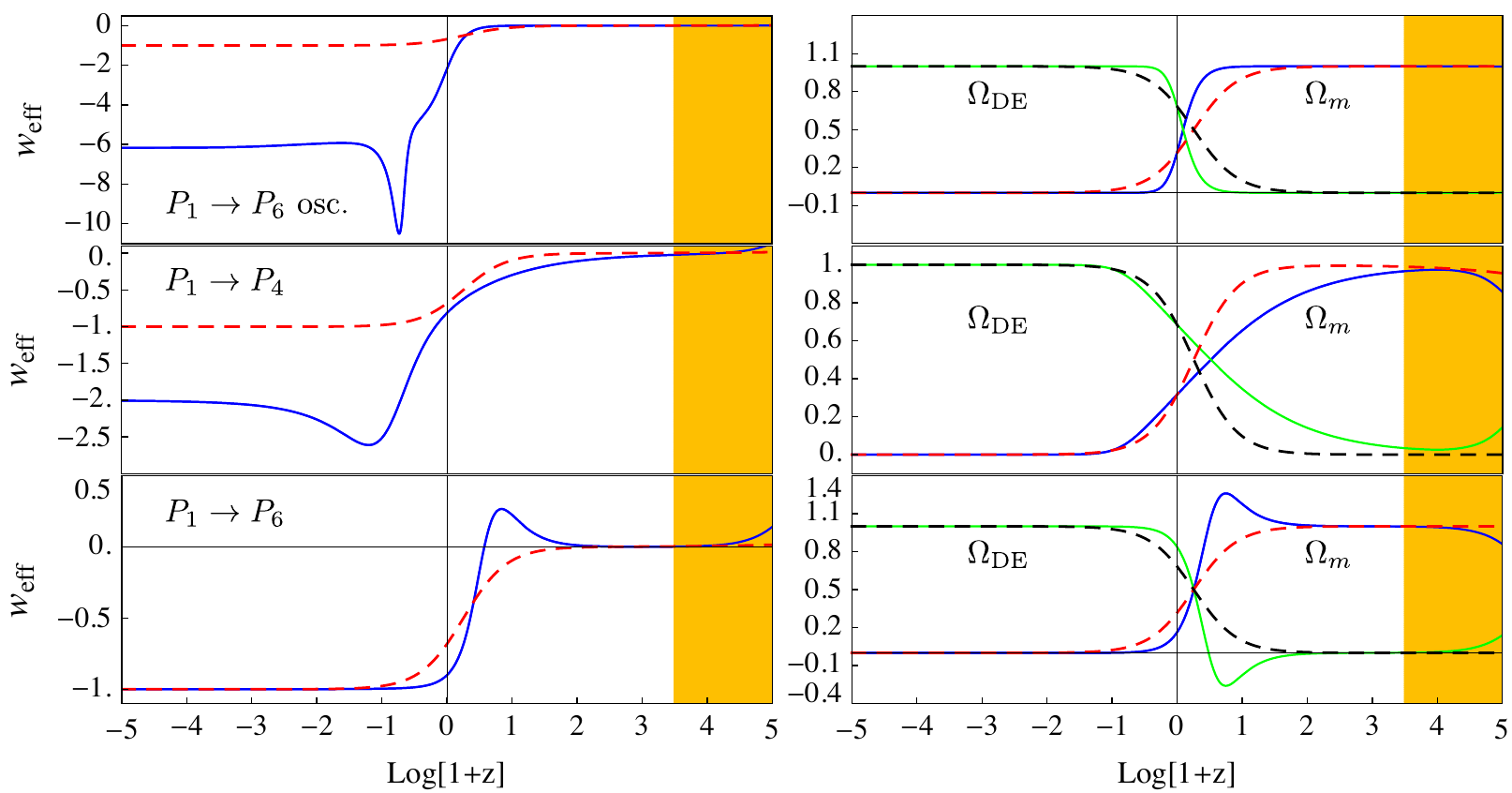}
\caption{The left panel shows the behavior of the effective equation of state for $\Lambda$CDM (red dashed line) and three different DE models (blue continuous line) corresponding to different types of trajectories identified in the \emph{first order} system and described in Sec.~\ref{Sub:FirstOrderNumerical}. The right panel shows the evolution of matter and dark energy densities for the $\Lambda$CDM model (respectively the red and black dashed lines) and the different DE models (respectively the blue and green solid lines). The yellow area represents the region in which we expect a non-negligible contribution from radiation which was not considered when constructing these numerical DE models.}
\label{Fig:FirstOrderCosmTrans}
\end{figure}
\underline{$P_1\rightarrow P_6$ transition with oscillations.}
We now investigate numerically a model which displays a $P_1 \rightarrow P_6$ transition with oscillations (Fig.~\ref{Fig:PhaseSpace}b). We obtain this behavior by setting $\alpha_1= 2.4$ and $\lambda_0=1.3$.
This time we impose initial conditions such that $\Omega_m=1$ at high redshift (i.e. $z=1000$), to evolve the system more into the future than in the past. Doing so, we avoid the dominance of the unstable point $P_2$ and are able to show a richer set of trajectories in the phase space plot. 
The most interesting family of trajectories corresponds to trajectories that either start at $P_2$ or infinity at early times, then pass close to $P_1$ crossing the $\Omega_m=1$ plane and then move close to $P_4$, and start circling toward $P_6$. The background cosmology of one of such trajectories is shown in Fig.~\ref{Fig:FirstOrderCosmTrans}.

\underline{$P_1\rightarrow P_4$ transition.}
In order to numerically recover a model which displays a $P_1 \rightarrow P_4$ transition, we choose $\alpha_1= -1/2$ and $\lambda_0=0$. 
The points $P_1$ and $P_2$ exhibit basically the same behavior as in the previous models, however for the chosen values of $\alpha_1,\lambda_0$ both $P_3$ and $P_4$  play the role of a dark energy attractor, with different $w_{\rm eff}$. 
This time we impose initial conditions to match the matter density today. In Fig.~\ref{Fig:PhaseSpace}c we can see as a result that we obtain two different types of trajectories that go from $P_1$ to $P_4$. The first set departs from $P_2$ and, after passing close to the matter saddle point $P_1$, go to the dark energy attractor $P_4$. The second one starts at infinity, then passes close to $P_1$ and eventually moves towards $P_4$. 
We plot the cosmological behavior of a trajectory that undergoes this transition in Fig.~\ref{Fig:FirstOrderCosmTrans}.

\underline{$P_1\rightarrow P_6$ transition.}
The last transition we  want to discuss corresponds is the $P_1 \rightarrow P_6$. In order to obtain trajectories with this behavior we set $\alpha_1= -2$ and $\lambda_0=-2$ and impose appropriate initial conditions in order to have equivalence between dark matter and dark energy density at the same redshift as the Planck best fit $\Lambda$CDM model~\cite{Ade:2013zuv}.
As we can see from the resulting phase space plot in Fig.~\ref{Fig:PhaseSpace}d, the system displays a clear transition from $P_1$ to $P_6$ for the trajectories that start close to $P_2$. In Fig.~\ref{Fig:FirstOrderCosmTrans} we show the cosmological behavior of one of these trajectories.\\
The selected values for $\alpha_1, \lambda_0$, allow also different types of trajectories, as can be read off Fig.~\ref{Fig:FirstOrderRegions}. In particular we can recognize two sets of trajectories that show a $P_1$ to $P_4$ transition. The first set of trajectories starts in $P_2$ and move toward $P_1$, but are then deviated towards $P_4$ instead of $P_6$. The second set of trajectories starts at infinity, approaches $P_1$ and then moves towards $P_4$.
Noticeably in the phase space plot in consideration (Fig.~\ref{Fig:PhaseSpace}d), one can observe non-trivial heteroclinic orbits joining $P_1$ and $P_4$, $P_4$ to $P_3$ and $P_6$ to $P_5$.

In summary, from the numerical investigation of the different transitions, we have learned that in general trajectories that undergo a $P_1\rightarrow P_3$ transition are those that closely resemble the $\Lambda$CDM cosmology. Models involving other transitions, such as $P_1\rightarrow P_4$ or $P_1\rightarrow P_6$, display trajectories that are quite different from the $\Lambda$CDM one, but still can give viable cosmologies as can be noticed in Fig.~\ref{Fig:FirstOrderCosmTrans}.

\subsection{Second order analysis}
\label{Secsecond}
We now proceed to the second order by allowing both $\alpha_0$ and $\alpha_1$ to vary, while fixing $\alpha_2$ and $\lambda_0$ to constant. The models under consideration will then be characterized by
\begin{align}
& \ddot{\Omega}(t)=\ddot{\Omega}_0 a^{-\alpha_2} \,, \hspace{1cm} c(t)-\Lambda(t)=(c-\Lambda)_0 a^{-\lambda_0}.
\end{align}
As it can be seen from~(\ref{syst}), $\alpha_2$ is the first of the $\alpha's$ that does not enter the core equations~(\ref{systx})-(\ref{systu}); it is therefore from this order on, that we start to observe some of the effects of the recursive nature of Eqs.~(\ref{systalpha}).
As we will shortly show, the majority of the critical points for the second order system are just trivial extensions of the critical points of the first order case; they replicate the values for the coordinates $\{x_c,y_c,u_c,\alpha_{0,c}\}$ and come in two copies distinguished by the value of $\alpha_1$, being it equal or different from zero. The latter difference reflects into a different dynamics for $\Omega(t)$, which can be richer for the points with $\alpha_1\neq 0$ . To highlight this splitting of the points, we shall label with the subscript $a$ the duplicates of the first order critical points that have $\alpha_1=0$, and with $b$ the duplicates that have ($\alpha_1\neq 0$). This splitting trend will become regular from the next order on and it will help us in Sec.~\ref{recursive} for  the classification of the points at a generic order $N$.\\
The  details of all the critical points and their stability are shown in Table~\ref{tabsecondorder} in Appendix~\ref{apsecond}. 
In what follows we briefly comment on the characteristics of the cosmologically interesting points.
\begin{itemize}[leftmargin=*]
\item\underline{\textsl{Matter points}}\\
There are two critical points that are matter dominated with $w_{\rm eff}=0$ and both of them represents the extension to one higher dimension of the $P_1$ critical point found in the first order analysis. Their coordinates and the eigenvalues of the linearized system are:
\begin{subequations}
\begin{align}
& P_{1a}\equiv\l(0,0,0,0,0\r) & & \mu_1=-3, \,\, \mu_2= -1, \,\, \mu_3=\f{3}{2}, \,\, \mu_4=\f{3}{2}-\alpha_2, \,\, \mu_5=3-\lambda_0. \\
& P_{1b}\equiv\l(0,0,0,0,\alpha_2-\f{3}{2}\r) & & \mu_1=-3, \,\, \mu_2= -1, \,\, \mu_3=3-\alpha_2, \,\, \mu_4=-\f{3}{2}+\alpha_2, \,\, \mu_5=3-\lambda_0.
\end{align}
\end{subequations}
The first one, $P_{1a}$, is a viable saddle point for $\lambda_0\neq3 \wedge \alpha_2 \neq \f{3}{2}$ while the second one, $P_{1b}$, is a saddle for $\lambda_0\neq0 \wedge \alpha_2 \neq \f{3}{2} \wedge \alpha_2 \neq 3$. 
As we can notice the stability requirements are quite mild if compared to the constraints that we found at the previous orders.
As a result the vast majority of second order models will have two cosmologically viable matter configurations distinguished by the behavior of $\Omega(t)$. 
When passing close to $P_{1a}$ models will be characterized by $\ddot{\Omega} \ll \dot{\Omega} \ll \Omega$ which means that the coupling to matter will be frozen at a certain value until the model moves toward dark energy domination. 
On the other hand the second configuration corresponds to a matter era in which $\Omega(t)$ has a non-trivial dynamics.
\begin{figure}[t]
\centering
\includegraphics[width=17.5cm]{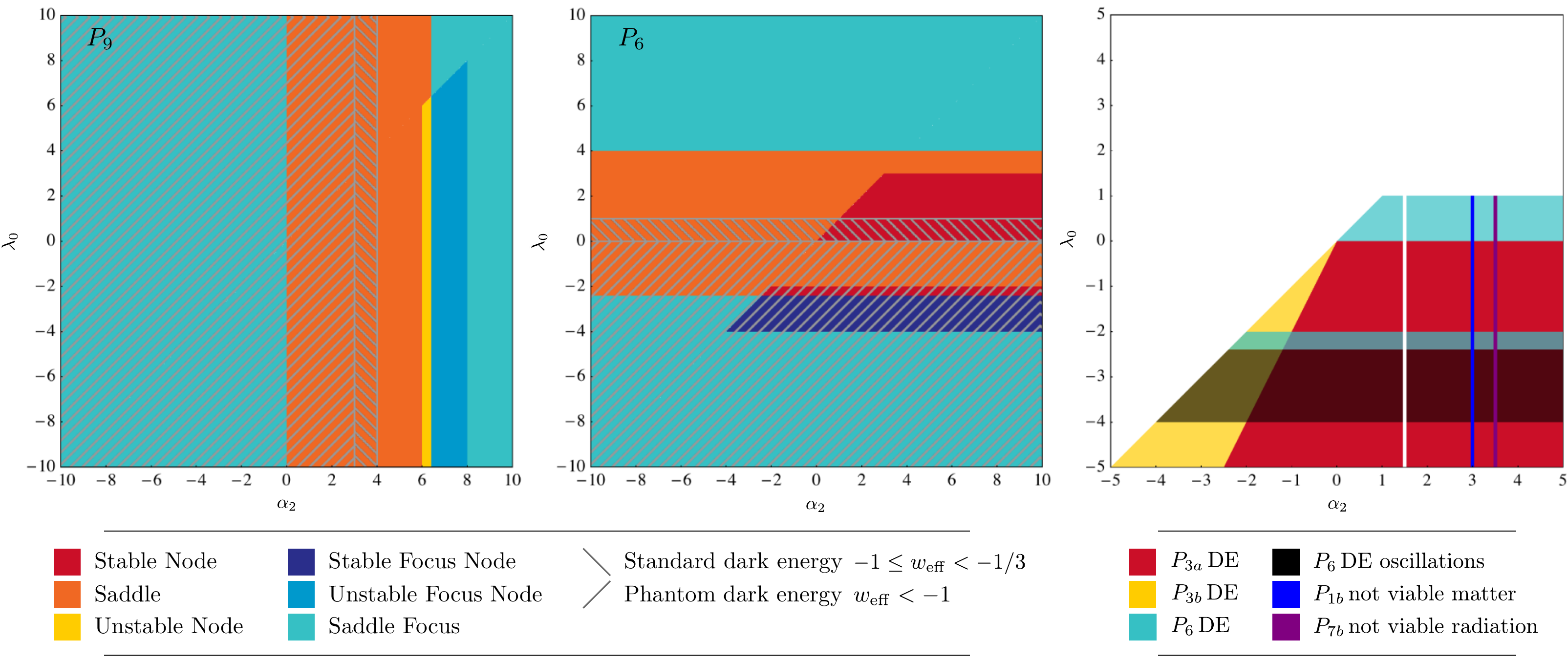}
\caption{The left panel shows the results of the stability analysis of the \emph{second order} system around $P_9$ (see Appendix~\ref{apsecond}). The panel at the center illustrates the stability of the system around $P_6$.
The right panel shows the combined results of the \emph{second order} analysis. Regions in the $(\alpha_2 , \lambda_0)$ plane which allow the different transitions discussed in Sec.~\ref{Secsecond} are shown in different colors.}
\label{Fig:SecondOrderRegions}
\end{figure}
\item\underline{\textsl{Stiff-matter points}}\\
There are two $P_2$-like critical points with a stiff matter equation of state, $w_{\rm eff}=1$:
\begin{subequations}
\begin{align}
& P_{2a}\equiv \l(1,0,0,0,0\r) & & \mu_1=2, \,\, \mu_2=3 , \,\, \mu_3=3, \,\, \mu_4=3-\alpha_2, \,\, \mu_5=6-\lambda_0.  \\
& P_{2b} \equiv \l(1,0,0,0,-3+\alpha_2\r) & & \mu_1=2, \,\, \mu_2=3 , \,\, \mu_3=6-\alpha_2, \,\, \mu_4=\alpha_2-3, \,\, \mu_5=6-\lambda_0.
\end{align}
\end{subequations}
Their unstable configuration, which might be relevant for the early stages of the Universe, can be obtained for  $\alpha_2 < 3 \wedge \lambda_0 < 6$ in the case of $P_{2a}$, and for $P_{2b}$ is $3 < \alpha_2 < 6 \wedge \lambda_0 < 6$ in the case of $P_1$. Again the two realizations of this point correspond to different behaviors of the conformal coupling $\Omega(t)$.
\item\underline{\textsl{Dark energy points}}\\
We have also two DE dominated points from the splitting of the first order $P_3$ point:
\begin{subequations}
\begin{align}
& P_{3a} \equiv \l(\frac{\lambda_0}{6},1-\frac{\lambda_0}{6},0,0,0\r) & & \mu_1=\lambda_0-6, \,\, \mu_2=\lambda_0-4 , \,\, \mu_3=\lambda_0-3, \,\, \mu_4=\f{\lambda_0}{2}, \,\, \mu_5=\f{1}{2}(\lambda_0-2\alpha_2). \\
& P_{3b} \equiv \l(\frac{\lambda_0}{6},1-\frac{\lambda_0}{6},0,0,\alpha_2-\f{\lambda_0}{2}\r) & & \mu_1=\alpha_2-\f{\lambda_0}{2}, \,\, \mu_2=\lambda_0-6 , \,\, \mu_3=\lambda_0-3, \,\, \mu_4=\lambda_0-4, \,\, \mu_5=\lambda_0-\alpha_2 .
\end{align} 
\end{subequations}
They both have $w_{\rm eff}=-1+ \lambda_0 / 3$ and are cosmologically viable late time DE attractors respectively for $(\alpha_2\geq0 \wedge \lambda_0<0) \vee (\alpha_2<0 \wedge \lambda_0<2\alpha_2)$ and for $\alpha_2<0 \wedge \lambda_0>2\alpha_2 \wedge \lambda_0<\alpha_2 $.\\
The other viable DE attractor is the second order equivalent of the dark energy dominated $P_6$~(\ref{Eq:FirstOrderP6}):
\begin{align}
& P_6\equiv \l(\f{\lambda_0}{2},1+\f{\lambda_0}{2},0 ,-\lambda_0,0\r) & & \mu_1=\lambda_0-4,  \,\,   \,\, \mu_2=\lambda_0-\alpha_2, \,\, \mu_3=-3-\f{3}{4}\lambda_0+\f{1}{4}\sqrt{3}\sqrt{48+8\lambda_0-5\lambda_0^2},\nonumber\\
& & & \mu_4=\lambda_0-3, \,\, \mu_5=-3-\f{3}{4}\lambda_0-\f{1}{4}\sqrt{3}\sqrt{48+8\lambda_0-5\lambda_0^2},  
\end{align}
which is an accelerated attractor with a viable equation of state for $(\alpha_2>1 \wedge 0<\lambda_0<1) \vee (0<\alpha_2\leq 1 \wedge 0<\lambda_0<\alpha_2)$.
From the full stability graphical analysis, reported in Fig.~\ref{Fig:SecondOrderRegions}, we can notice that this point can be an accelerated attractor for a wider range of $(\alpha_2,\lambda_0)$, however for some intervals it would have $w_{\rm eff}<-2$, which is a value already excluded by experiments, e.g.~\cite{Ade:2013zuv,Rest:2013bya}, and hence we have considered a more conservative region.
\item\underline{\textsl{Radiation points}}\\
Two radiation dominated critical points are provided by the splitting of the first order point $P_7$:
\begin{subequations}
\begin{align}
& P_{7a}\equiv \l(0,0,1,0,0\r)& & \mu_1=-2, \,\, \mu_2=1 , \,\, \mu_3=2, \,\, \mu_4=\f{3}{2}-\alpha_2, \,\, \mu_5=4-\lambda_0.  \\
& P_{7b}\equiv \l(0,0,1,0,-\f{3}{2}+\alpha_2\r) & & \mu_1=-2, \,\, \mu_2=1 , \,\, \mu_3=\f{7}{2}-\alpha_2, \,\, \mu_4=\alpha_2-\f{3}{2}, \,\, \mu_5=4-\lambda_0.
\end{align}
\end{subequations}
They are a saddle respectively for $\lambda_0\neq4 \wedge \alpha_2 \neq \f{3}{2}$ and $\lambda_0 \neq 4 \wedge \alpha_2 \neq \f{7}{2} \wedge \alpha_2 \neq \f{3}{2}$. 
A viable radiation era can also be provided by $P_{10}$ (see Table~\ref{tabsecondorder}) which is a radiation-DE scaling critical point.
The stability analysis of this critical point is too complicated to be shown because of the complexity of its eigenvalues; nevertheless we can deduce the stability conditions for the configurations of cosmological interest. For instance for $\alpha_2=\frac{7}{2}$ this point supplies a good radiation dominated point which is a saddle if $\lambda_0 \neq 4$. We cannot instead identify a region of $(\alpha_2,\lambda_0)$ where this point would provide a viable DE candidate.
\end{itemize}
 Combining the above results, we can see that for the second order system there is a wide variety of possible transitions between viable critical points that will give rise to a working cosmological model. This is somewhat expected given that we are moving up the $\alpha$ channel and allowing more and more general behaviors of the function $\Omega(t)$.
The combined results of the second order dynamical analysis are shown in Fig.~\ref{Fig:SecondOrderRegions}. 
In general the stability requirements for a viable radiation and matter era are much less stringent than those for the first order system. Indeed, except for a discrete set of values of $\alpha_2,\lambda_0$, generally  there are two points that can give a radiation era, i.e. $P_{7a}$ or $P_{7b}$,  as well as two points that can provide a matter era, i.e. $P_{1a}$ or $P_{1b}$. The values of $\alpha_2$ that do not allow either a viable matter or radiation critical point  are shown in Fig.~\ref{Fig:SecondOrderRegions} as, respectively, straight blue and purple lines. A stronger selection of viable regions in the $(\alpha_2,\lambda_0)$ plane is imposed by requiring that the possible DE points, $P_{3a}, P_{3b}, P_{6}$, have a proper cosmology and stability. 

\subsection{${\rm N}^{\rm th}$ order analysis: exploiting the recursive nature of the system}
\label{recursive}
In the previous Sections we performed a dynamical analysis of the system~(\ref{syst}) cutting the hierarchy of equations~(\ref{systalpha}) at increasingly higher orders, up to the second, while keeping $\lambda_0$ constant. 
At each order we determined the critical points, their stability and cosmological features.
The reason for treating separately the zero, first and second order is twofold. First, it allows us to study gradually more and more general models, recognizing at each order some characteristic features and cosmological viability conditions. Second, since $\alpha_2$ is the first of the $\alpha's$ not to enter the core equations~(\ref{systx})-(\ref{systu}), we expect that from the third order up the system will display a regular pattern in the critical points that reflects the recursive structure of the equations~(\ref{systalpha}). We saw glimpses of this pattern already in the second order system in~\ref{Secsecond}, but it is not until we have $N\geq3$ that it displays fully. We will now exploit this feature to reconstruct the dynamical properties of the system at any given order $N\geq3$, building on the findings of the lower order analyses. We neglect radiation  for simplicity (our results can be easily extended to include it), so we are left with an $N+2$ dimensional system for the variables $\{x,y,\alpha_0,\alpha_1,\dots,\alpha_{N-1}\}$. When writing the coordinates of the critical points we use the general structure $(x_c,y_c,\alpha_{0,c}, \alpha_{1,c}, \alpha_{n,c})$, with $n=2,..,N-1$, which allows us to treat separately $\alpha_0,\alpha_1$ from $\alpha_n$ with $n\geq2$, given that the former enter the core equations~(\ref{systx})-(\ref{systu}) and do not obey the general rules that we are about to derive.

By looking at system~(\ref{syst}), one notices that the set of variables $\{x,y,\alpha_0\}$ depends on the $\alpha_n$, $n\geq2$, only through $\alpha_1$. We can therefore use $\alpha_1$ as a pivot variable and split the original system into two blocks: the block of equations~(\ref{systx}),~(\ref{systy}),~(\ref{systalpha}) with $n=1$ and the block of equations~(\ref{systalpha}) with $n\geq2$. We start by solving the equations of the first block, and determine solutions for ($x_c, y_c, \alpha_{0,c}$) as functions of $\alpha_1$. We then turn to the second block and notice that one can generally distinguish two cases: those characterized by $\alpha_{1,c}=0$ and those with $\alpha_{1,c}\neq 0$. In the former case, the two blocks are  independent, while in the latter all the coordinates of the critical points will be affected by the equations of the second block. The general structure of the points for which $\alpha_1=0$ can then be recovered as follows. One starts solving the first block of equations, which can be done quite straightforwardly, to determine $\{x_c,y_c,\alpha_{0,c}\}$. Then one turns the attention to the second block, with $n\geq 3$ since $\alpha_{1,c}=0$, and finds that there are three types of general solutions for this block: one in which all $\alpha_{n,c}=0$, the second where all $\alpha_{n,c}\neq 0$ and the last case in which there will be different combinations of $\alpha's$ equal or not to zero (hence the name \textit{combinations} in what follows). A combination is specified by the location of all the zero terms; once these are given, the values of the $\alpha's\neq0$ are uniquely determined and can be reconstructed, after some lengthy algebra, solving the corresponding equations. Let us illustrate the general rules for the specific expressions of the non-zero $\alpha's$, by using the following representative \textit{combination}:
\begin{equation}\label{comb}
\alpha_{n,c}\equiv (\,\underbrace{ 0, \dots,0 }_{\text{block}\,=\,0} \,,\, \underbrace{\alpha_{q,c}, \dots\alpha_{j,c}\dots, \alpha_{s,c}}_{\text{block}\,\neq\, 0, \,\, j=q,...s} \,,\, \underbrace{ 0, \dots ,0 }_{\text{block}\,=\,0} \,,\, \underbrace{\dots\alpha_{j,c}\dots}_{\text{block}\,\neq\,0}\,,\, \underbrace{0,\dots,0}_{\text{block}\,=\,0}\,,\, \underbrace{\alpha_{k,c},...\alpha_{l,c}...,\alpha_{N-1,c}}_{\text{block}\,\neq\,0, \,\, l=k,...,N-1}\,).
\end{equation}
The  elements in the non-zero blocks which are followed by a zero block have:
\begin{equation} \label{rule1}
\alpha_{j,c}=(s+1-j)\frac{\dot{H}}{H^2},
\end{equation}
where $q\leq j\leq s$, with $\alpha_q$ being the first non-zero term in the block and $\alpha_s$ the last one. The particular \textit{combination} shown in~(\ref{comb}) ends with a non-zero block;  the elements of such a block obeys the following specific rule:
\begin{equation}\label{rule2}
\alpha_{l,c}= \alpha_{N}+(N-l)\frac{\dot{H}}{H^2}, 
\end{equation}
where $k\leq l\leq N-1$, with $\alpha_k$ being the first non-zero term in the block. Every time we substitute into~(\ref{rule1}) and~(\ref{rule2}) the specific value of $\dot{H}/H^2(x_c,y_c,\alpha_{0,c},\alpha_{1,c})$ that corresponds to the point in consideration.

The solutions for which the variable $\alpha_{1}$ assumes a non-zero value are a little trickier to treat as the components ($x_c,y_c,\alpha_{0,c}$) of the critical points will be affected by the equations of the second block,  we find that also in this case the critical points can generally be separated in the three above cases based on the structure of the $\alpha_n$, $n\geq 2$, block for which the general rules~(\ref{rule1})-(\ref{rule2}) still apply.

Using the above technique we are able to reconstruct all the critical points of system~(\ref{syst}) at a given order $N$. In particular, we find that they can be organized in families characterized by  the same cosmological behavior. These families, in most of the cases, can be directly connected to the critical points that we have analyzed in the previous Sections, as expected because of the structure of our system and its invariant manifolds (as mentioned at the end of the introductory part of Sec.~\ref{Sec3}). Therefore one can identify the main critical points of cosmological interest, or in other words get a good sense of the cosmologies encoded in the EFT formalism, already at the lower orders. Going to higher orders allows to analyze more and more general models.

In what follows we describe only the families of critical points that allow for a viable cosmology, leaving the discussion of the remaining critical points for Appendix~\ref{apgeneral}. We generally indicate with $s$ the position of the last term in a non-zero block within the combination, and with $k$ the position of the first non-zero term in the last  non-zero block that, for some cases, closes the combination.
\begin{itemize}[leftmargin=*]
\item\underline{\textsl{Matter points:}} \\
This family includes $2^{N-1}$, $P_1$-like, critical points characterized by a well defined cosmology ($\Omega_m=1$):
\begin{subequations}
\begin{eqnarray}
&&P_{1a}\equiv(0, 0, 0, 0, \alpha_{n,c}=0), \\
&&P_{1b}\equiv\left(0, 0, 0, \alpha_{N}-\frac{3}{2}\l(N-1\r), \alpha_{n,c}=\alpha_{N}-\frac{3}{2}(N-n)\right),    \\
&&P_{1c}\equiv(0, 0, 0, {\rm combinations}).
\end{eqnarray}
\end{subequations}
The latter point includes all ($2^{N-1}-2$) possible combinations constructed via Eqs.~(\ref{rule1}) and~(\ref{rule2}) with $\dot{H}/H^2=-1$.
All critical points correspond to matter domination, therefore, instead of performing the full stability analysis, we simply determine the intervals for which they are saddle points. 
The eigenvalues of the linearized system around $P_{1a}$ and $P_{1b}$ are:
\begin{subequations}
\begin{align}
& P_{1a}: & & \mu_1=-3, \,\, \mu_2=\f{3}{2}-\alpha_{N}, \,\, \mu_3=3-\lambda_0, \,\, \mu_4= \dots = \mu_{N-1} = \f{3}{2},\\
& P_{1b}: & & \mu_1= -3, \,\, \mu_2=\f{3}{2}N-\alpha_{N}, \,\, \mu_3=3-\lambda_0 , \,\,  \mu_4= \dots = \mu_{N-1} = \alpha_{N}-\f{3}{2}(N-h),
\end{align}
\end{subequations}
where $h=1,..,N-1$.
As we can see these points have only one possible stability configuration having two eigenvalues of opposite sign, therefore as long as they are hyperbolic they are saddles.
The first one is hyperbolic if $\lambda_0\neq3$ and $\alpha_{N}\neq 3/2$ while for the second one we should have $\alpha_{N} \neq \f{3}{2}\l( N-h\r) $,  $\alpha_{N} \neq \f{3}{2}N $ and $\lambda_0 \neq 3$. As for the last sub-family of critical points, $P_{1c}$, the analysis of the eigenvalues reveals that this is a set of saddle points regardless of the particular combination as for each combination at least two eigenvalues have opposite sign.  Despite the complexity of the structure of the combinations, we are able to determine that all of them are hyperbolic if: $\lambda_0\neq 3$ and $\alpha_N\neq \f{3}{2}(N-h) \,\,\, \text{with} \,\,\, h=1,\dots,N-1$.
\item\underline{\textsl{Stiff-matter points:}}
\begin{subequations}
\begin{eqnarray}
&&P_{2a}\equiv(1, 0, 0, 0, \alpha_{n,c}=0 ),\\
&&P_{2b}\equiv\left(1, 0, 0, \alpha_{N}-3(N-1), \alpha_{n,c}=\alpha_{N}-3(N-n)\right),\\
&&P_{2c}\equiv(1, 0, 0, {\rm combinations}),\,\, \alpha_{j,c}=-3(s+1-j)\,\,,\,\,\alpha_{l,c}=\alpha_{N}-3(N-l).
\end{eqnarray}
\end{subequations}
The points in this family have $\Omega_{\rm DE}=1$ and $w_{\rm eff}=1$, therefore representing a set of $2^{N-1}$ stiff-matter critical points. 
The structure and the cosmology of these critical points suggest a similarity with the $P_2$ critical point we analyzed in the previous Sections. These critical points could be of interest in the early stages of the Universe as unstable critical points~\cite{Copeland:1997et}, which is the only configuration we analyze in what follows.
The first two critical points have eigenvalues:
\begin{subequations}
\begin{align}
& P_{2a}: & & \mu_1=3-\alpha_{N}, \,\, \mu_2=6-\lambda_0, \,\, \mu_3=\mu_4=\dots= \mu_{N-1} =3,\\
& P_{2b}: & & \mu_1=3,\,\, \mu_2=3N-\alpha_{N}, \,\, \mu_3={6-\lambda_0},\,\, \mu_4=\dots= \mu_{N-1} = \alpha_{N}-\frac{3}{2}\l(N-1-h\r),
\end{align}
\end{subequations}
where $h=1,..,N-1$. The first critical point is unstable for $\alpha_{N}<3 \wedge \lambda_0<6$ while the unstable configuration of the second one is obtained if $ 3/2 \l(N-2 \r) < \alpha_{N} <  3 N \wedge \lambda_0<6$.
For the last sub-family, $P_{2c}$,  there is only one combination which shows an unstable configuration and it is the one with $\alpha_1=0$ and $\alpha_{n,c}\neq0$ for $n=2,\dots,N-1$ which is unstable if $ \lambda_0<6  \wedge 3<\alpha_{N}<3(N-1)$. Most of the other combinations are saddle points.

\item\underline{\textsl{Dark Energy points:}}
\begin{subequations}
\begin{eqnarray}
&&P_{3a}\equiv\left( \f{\lambda_0}{6}, 1-\f{\lambda_0}{6}, 0, \alpha_{n,c}=0 \right),\\
&&P_{3b}\equiv\left( \f{\lambda_0}{6}, 1-\f{\lambda_0}{6}, 0,\alpha_{N}-\frac{\lambda_0}{2}\l(N-1\r),  \alpha_{n,c}=\alpha_{N}-\frac{\lambda_0}{2}\l(N-n\r)\right),\\
&&P_{3c}\equiv\left( \f{\lambda_0}{6}, 1-\f{\lambda_0}{6},0, {\rm combinations}\right),\,\, \alpha_{j,c}=-(s+1-j)\f{\lambda_0}{2} \,\,,\,\, \alpha_{l,c}=\alpha_{N}-\f{\lambda_0}{2}(N-l).
\end{eqnarray}
\end{subequations}
This family corresponds to a set of $2^{N-1}$ DE dominated critical points with $\Omega_{\rm DE}=1$ and $w_{\rm eff}=\f{\lambda_0}{3}-1$. From the structure of these points we can immediately recognize a similarity with the $P_3$ critical point analyzed in the previous Sections. We are interested in the stable configuration for this family.
The eigenvalues of the system around the first two points are:
\begin{subequations}
\begin{align}
& P_{3a}: & & \mu_1=\lambda_0-6, \,\, \mu_2= \lambda_0-3, \,\, \mu_3= \f{\lambda_0}{2}-\alpha_{N}, \,\, \mu_4=\dots= \mu_{N-1}=\f{\lambda_0}{2}, \\
& P_{3b}: & & \mu_1=\lambda_0-6,\,\, \mu_2= \lambda_0-3, \,\, \mu_3=\f{\lambda_0}{2} , \,\,  \mu_4=\dots= \mu_{N-1}= \f{\lambda_0}{2}+\alpha_{N}-\frac{3}{2}(N-h+1),
\end{align}
\end{subequations}
where  $h=1,\dots,N-1$. The stability analysis reveals that $P_{3a}$ is a stable accelerated attractor if $(\alpha_{N}>0 \wedge \lambda_0<0) \vee (\lambda_0<2\alpha_{N}\wedge \alpha_{N}\leq 0)$ while $P_{3b}$ displays this cosmological behavior if $(\lambda_0<0 \wedge \alpha_N\leq3) \vee (\alpha_{N}>3 \wedge \lambda_0<6-2\alpha_N)$. The last sub-family $P_{3c}$ does not contain any stable solution, and as a consequence will be not further considered.
\end{itemize}

\noindent The points discussed above represent all the hyperbolic, cosmologically viable, critical points of the system~(\ref{syst}) at a given order $N\geq3$ (with $\lambda_0={\rm constant}$).
Since we neglected radiation, the families of critical points which are of cosmological interest and that can be used to construct transitions from a matter era to a DE one are, respectively, the $P_1$-like and $P_3$-like family. Each family contains several critical points, therefore there are many possible specific transitions; in particular at a given order $N$, there are  $2^{N-1}$ matter points and $2$ DE points.
Analogously to what happens in the second order case, the intervals of cosmological viability for ($\alpha_N, \lambda_0$)
are strongly influenced by the stability requirements of the DE points, while the requirements for a good matter era are significantly easier to satisfy, and only exclude some values.
Once one selects the values of ($\alpha_{N}$, $\lambda_0$) according to the intervals reported above, the trajectories of the dynamical system will generally start at early times in the neighborhood of a $P_2$-like unstable node then approach a $P_1$-like matter point,  finally moving away from it heading towards a $P_3$-like de-Sitter attractor. Different trajectories will correspond to different behaviors of the EFT functions $\Omega(t)$ and $c(t)$. Let us conclude pointing out that viable transitions have $\lambda_0<0$, which implies that $c(t)-\Lambda(t)$ will be a growing function of time for all viable trajectories at the $N^{\rm th}$ order. 

\section{Conclusions}\label{conclusions}
In this paper we performed a thorough dynamical analysis  of the background cosmology within the effective field theory of dark energy formalism~\cite{Gubitosi:2012hu,Bloomfield:2012ff}. In particular we  investigated general conditions of cosmological compatibility for the three functions of time that describe the background dynamics in this formalism (EFT functions). While the system of equations is underdetermined, we identified a set of variables that allows one to transform it into an infinite-dimensional system characterized by an important recursive structure. We  then studied several autonomous cases of increasingly higher dimension corresponding to more and more general models of dark energy and modified gravity within the EFT framework. Furthermore, exploiting the recursive nature of the full system of equations, as well as our findings at the lower orders, we drew some general conclusions on its dynamics and cosmological behavior. 

Our set of dynamical variables contains two infinite series of variables $\alpha_n$ and $\lambda_m$, defined as ratios of subsequent derivatives of, respectively, the EFT functions $\Omega$ and $c-\Lambda$,~(\ref{variables}). These variables are such that their corresponding equations assume a hierarchical structure, that we exploit throughout the paper. One can truncate these series at any desired order, and study the corresponding autonomous system. We focused on the $\alpha$ channel, keeping always $\lambda_0$ constant.  In other words, we focused on the class of models for which $c-\Lambda$ is a power law in the scale factor, while the conformal factor $\Omega$ can be increasingly general as we go up with the order. Alternatively one could fix $\Omega$ to a constant and open the $\lambda$ channel, which would correspond to exploring all minimally coupled models of DE. Finally, one could work with both channels and, for instance, explore, within this parametrized framework the full class of Horndeski theories~\cite{Horndeski:1974wa}. While we leave the former for future work, we want to stress that the machinery set up in this paper is general and easily applicable to the other cases mentioned above. 

As we showed in~(\ref{Taylor_exp}), our set up allows us to find a general expression for $\Omega$ consisting, at a given order $N$, of a Taylor expansion of order $N-1$ in time and the corresponding remainder that is parametrized in terms of $\alpha_N$. Since we include the remainder, increasing the order of the analysis is not a matter of increasing the precision of the Taylor expansion but rather it allows the investigation of a wider class of models of dark energy and modified gravity with the most diverse coupling.  An analogous argument could be repeated for $c-\Lambda$. 

Focusing on the $\alpha$ variables, while keeping $\lambda_0$ constant, we analyzed the system at increasingly higher order. At each order we found the critical points and analyzed their stability and cosmological nature, determining regions in the plane ($\alpha_N$, $\lambda_0$) which allow for viable cosmological trajectories. 
The simplest case we analyzed was the zero-th order one, obtained setting $\alpha_0$ to constant. It corresponds to a power law behavior for the EFT functions. After finding the critical points, we performed a stability analysis and determined the cosmology of each point as function of $\alpha_0,\lambda_0$.  The general result of the zero-th order analysis is that viable cosmological models can be recovered setting $\alpha_0=0$ and $\lambda_0 \approx 0$ and there is really only one viable transition between cosmologically interesting critical points. Given that $\alpha_0=0$, the corresponding models will be characterized by a constant conformal factor $\Omega$, which is just a rescaling of the Planck mass. In Sec.~\ref{quintessence} we showed how these findings, projected onto models of quintessence, imply that a quintessence model with a potential which is a power law in the scale factor, cannot have a power law behavior for $\Omega$ and therefore, at this order is forced to be minimally coupled. We then proceeded with the analysis of the first and second order systems, finding, as expected, a richer set of cosmologies. We identified respectively the $(\alpha_1,\lambda_0)$ and $(\alpha_2,\lambda_0)$ regions which result in cosmologically compatible EFT functions.

At the second order we started to notice some reflections of the recursive nature of the equations for the $\alpha's$ . In particular, we found that the majority of the critical points for the second order system are just trivial extensions of the critical points of the first order case,  that come in two copies with similar cosmology but a different dynamics of the conformal factor $\Omega(t)$. The recursive nature of the dynamical system fully displays when $N\geq3$, which is part of the reason why we treated separately the zero, first and second order cases. In Sec.~\ref{recursive} we showed how to exploit this recursive feature to reconstruct the critical points, their stability and their corresponding cosmological dynamics at any given order $N\geq 3$. We identified regions in ($\alpha_N$, $\lambda_0$, $N$) space that allow compatible forms of the EFT functions; in particular, all viable models correspond to a function $c-\Lambda$ that grows in time.

Our methodology offers a general tool to perform the dynamical analysis of dark energy and modified gravity models within the EFT language.  In this paper we have used it to explore models with an increasingly more general conformal coupling; we leave  the analysis of other realizations for future work.
Finally, let us point out that in this paper we have chosen not to perform a full analysis of the scaling configurations, but rather focused on the two extreme cases for which either of the two components in the configuration has fractional energy density equal to unity. While we leave a thorough investigation of the scaling regime for future work, we expect that the scaling points that we found, especially at the order $N\geq 3$, will display a rich phenomenology of late-time scaling cosmologies that could provide a dynamical solution to the coincidence problem. 

We plan to apply our results to numerical investigations of the dynamics of linear perturbations within the model-independent framework of effective field theory of dark energy. Given the generality of the formalism, there is a high degree of freedom so that even after fixing the expansion history one is left with a completely undetermined function of time out of the three original EFT functions. As such, a designer approach that fixes the background cosmology (typically to $\Lambda$CDM) and uses the Friedmann equations to reconstruct the corresponding behavior of the EFT functions, may not be the optimal way to proceed. With our technique we are able to explore the cosmological dynamics of several forms of the EFT functions and determine general conditions of cosmological compatibilty at different order. This will help us in choosing appropriate ans$\ddot{\text{a}}$tze for the EFT background functions to input in numerical boltzmann codes that study the evolution of linear perturbations.

\section*{Acknowledgments}
We are grateful to Carlo Baccigalupi for his feedback on this work and for fruitful conversations, and to Jolyon Bloomfield, Tamara Grava, Stefano Luzzatto, Levon Pogosian, Daniele Vernieri, Shuang-Yong Zhou for useful discussions. NF acknowledges partial financial support from the European Research Council under the European Union Seventh 
Framework Programme (FP7/2007-2013) / ERC Grant Agreement n.~306425 ``Challenging General Relativity'' and from 
the Marie Curie Career Integration Grant LIMITSOFGR-2011-TPS Grant Agreement n.~303537. AS acknowledges support from a SISSA Excellence Grant, and partial support from the INFN-INDARK initiative.

\appendix
\section{Second Order Analysis continued}\label{apsecond}
In this Appendix we complete the analysis of the critical points of the second order system. In particular, all 
critical points (including those already discussed in Sec.~\ref{Secsecond}) and their stability analysis are reported in Table~\ref{tabsecondorder}; while  in the following we present the eigenvalues and discuss the cosmology of the points that were not considered in Sec.~\ref{Secsecond}.
\begin{table}[h!]
\centering
\footnotesize
\renewcommand\arraystretch{1.7}
\begin{tabular}{cp{4.5cm}p{6cm}*{3}{c}}
\hline
\hline
\centering{Point} & \centering{$\l[x_c,y_c,u_c,\alpha_{0,c},\alpha_{1,c}\r]$}& \centering{Stability} & $\Omega_{\rm DE}$ & $w_{\rm{eff}}$  \\
\hline
\hline
\centering{$P_{1a}$} & \centering{$\l[0,0,0,0,0\r]$} & \centering{{\bf Saddle point:} $\lambda_0 \neq 3\wedge \alpha_2 \neq \frac{3}{2}  $}   & 0 &  0         \\
\hline
 \centering{$P_{1b}$} & \centering{$\l[0,0,0,0,\alpha_2-\f{3}{2}\r]$} &  \centering{ {\bf Saddle point:} $ \lambda_0\neq3 \wedge \alpha_2 \neq \f{3}{2} \wedge \alpha_2 \neq 3$ }  & $0$& $0 $  \\
\hline
\centering{$P_{2a}$} & \centering{$\l[1,0,0,0,0\r]$} & \centering{ {\bf Unstable node:} $\alpha_2<3 \wedge \lambda_0<6$ \\  {\bf Saddle point:} otherwise } & 1 & 1  \\
\hline
\centering{$P_{2b}$} & \centering{$\l[1,0,0,0,-3+\alpha_2\r]$} & \centering{{\bf Unstable node:} $3<\alpha_2<6 \wedge \lambda_0<6 $ \\ {\bf Saddle point:} otherwise } & 1 & 1  \\
\hline
\centering{$P_{3a}$} & \centering{$\l[\frac{\lambda_0}{6},1-\frac{\lambda_0}{6},0,0,0\r]$}   & \centering{ {\bf Stable node:} $(\alpha_2\geq0 \wedge \lambda_0<0) \vee (\alpha_2<0 \wedge \lambda_0<2\alpha_2)$, \hspace{3cm}{\bf Unstable node:} $(\lambda_0>6  \wedge \alpha_2<3) \vee (\lambda_0>2 \alpha_2 \wedge \alpha_2\geq3)$,\hspace{3cm} {\bf Saddle point:} otherwise} & 1 & $\frac{1}{3}(\lambda_0-3)$  \\
\hline
\centering{$P_{3b}$} & \centering{$\l[\frac{\lambda_0}{6},1-\frac{\lambda_0}{6},0,0,\alpha_2-\f{\lambda_0}{2}\r]$}   & \centering{ {\bf Stable node:} $ \alpha_2<0 \wedge 2 \alpha_2<\lambda_0 \wedge \lambda_0<\alpha_2$, \hspace{3cm}{\bf Unstable node:} $(\alpha_2\geq6 \wedge \alpha_2< \lambda_0 \wedge \lambda_0<2\alpha_2) \vee (\alpha_2>3 \wedge \lambda_0>6 \wedge \alpha_2<6 \wedge \lambda_0<2\alpha_2)$,\hspace{3cm} {\bf Saddle point:} otherwise} & 1 & $\frac{1}{3}(\lambda_0-3)$  \\
\hline
\centering{$P_{4a}$} & \centering{$\l[-1,0,0,2,0\r]$} &  \centering{ {\bf Stable node:} $\alpha_2>2 \wedge \lambda_0>-2$ \\ {\bf Saddle point:} otherwise}  & 1& $-\frac{7}{3}$  \\
 \hline
\centering{$P_{4b}$} &  \centering{$\l[-1,0,0,2,1+\f{\alpha_2}{2}\r]$} &  \centering{ {\bf Stable node:} $\alpha_2<-2 \wedge \lambda_0>\alpha_2$, \\ {\bf Unstable node:} $ \alpha_2>4\wedge \lambda_0<\alpha_2$ \\ {\bf Saddle point:} otherwise }  & $1$& $\f{1}{3}(-5+\alpha_2) $  \\
 \hline
\centering{$P_{5}$} &  \centering{$\l[-1,0,0,-3,0\r]$} &  \centering{ {\bf Stable node:} $\alpha_2>3 \wedge \lambda_0>3$, \\ {\bf Saddle point:} otherwise}  & $-4$& $1 $  \\
\hline
\centering{$P_{6}$} & \centering{$\l[\f{\lambda_0}{2},1+\f{\lambda_0}{2},0 ,-\lambda_0,0\r]$} &  \centering{ \bf See Fig.~\ref{Fig:SecondOrderRegions}}
 &  1 &  $-1+\f{2\lambda_0}{3}$  \\
 \hline
\centering{$P_{7a}$} &  \centering{$\l[0,0,1,0,0\r]$}&  \centering{{\bf Saddle point:} $\lambda_0 \neq 4 \wedge \alpha_2\neq\f{3}{2}  $}   & 0  & $\frac{1}{3} $         \\
\hline
\centering{$P_{7b}$} & \centering{$\l[0,0,1,0,-\f{3}{2}+\alpha_2\r]$}&  \centering{{\bf Saddle point:} $\lambda_0 \neq 4 \wedge \alpha_2\neq \frac{7}{2} \wedge \alpha_2 \neq \frac{3}{2}$}   & 0  & $\frac{1}{3} $         \\
\hline
\centering{$P_{8}$} & \centering{$\l[-4,0,9,-4,0\r]$}&  \centering{{\bf Saddle point:} $\lambda_0 \neq 4 \wedge \alpha_2\neq-\f{1}{2}  $}   & -8  & $\frac{5}{3} $         \\
\hline
\centering{$P_{9}$} & \centering{ $\l[\alpha_2-5,0,0,6-\alpha_2,3\r]$} &  \centering{ {\bf See Fig.~\ref{Fig:SecondOrderRegions}}} & $1$ & $-3+\frac{2\alpha_2}{3} $   \\
\hline
\centering{$P_{10}$} & \centering{$\l[-9 + 2 \sqrt{8 - 2 \alpha_2} + 2 \alpha_2,0,\r.$\\ $\l.8 - 2 \alpha_2,2 - 2 \sqrt{8 - 2 \alpha_2},3 - \sqrt{8 - 2 \alpha_2}\r]$} &  \centering{ {\bf See Sec.~\ref{Secsecond}} }   & $-7 + 2 \alpha_2$ & $\f{1}{3} \left(-1 + 2 \sqrt{8 - 2 \alpha_2}\right) $ \\
\hline
\centering{$P_{11}$} & \centering{$\l[-9 - 2 \sqrt{8 - 2 \alpha_2}+ 2 \alpha_2,0,\r.$\\ $\l.8 - 2 \alpha_2,2 + 2 \sqrt{8 - 2 \alpha_2},3 + \sqrt{8 - 2 \alpha_2}\r]$} &  \centering{ {\bf See Appendix~\ref{apsecond}} }  & $-7 + 2 \alpha_2$ & $\f{1}{3} \left(-1 - 2 \sqrt{8 - 2 \alpha_2}\right) $ \\
\hline
\hline
\end{tabular}
\renewcommand\arraystretch{1}
\caption{Hyperbolic critical points for the second order system with $\alpha_2,\lambda_0={\rm constant}$. Taking into account also the additional constraints $\Omega_{\rm m}\geq0$ and  $\Omega_{\rm r}\geq0$, the domain for last two critical points is $\mathcal{D}\equiv\left\{ \alpha_2< 4\mbox{,} \, \lambda_0 \in \mathbb{R} \r\}$, while all other points have $\mathcal{D}\equiv\left\{  \alpha_2 \mbox{,} \,\lambda_0 \in \mathbb{R} \right \}$.}\label{tabsecondorder}
\end{table}
\begin{itemize}[leftmargin=*]
\item\underline{\textsl{Phantom DE points}} \\
From the splitting of the first order point $P_4$, we have two critical points characterized by a phantom effective equation of state:
{\small\begin{subequations}
\begin{align}
& P_{4a}: & & \mu_1=-6, \,\, \mu_2=-5 , \,\, \mu_3=-3, \,\, \mu_4=-2-\alpha_2, \,\, \mu_5=-2-\lambda_0. \\
& P_{4b}: & & \mu_1=\f{1}{2}(\alpha_2-4), \,\, \mu_2=\alpha_2-4 , \,\, \mu_3=\alpha_2-3, \,\, \mu_4=\alpha_2+2, \,\, \mu_5=\alpha_2-\lambda_0.
\end{align}
\end{subequations}}
The first one has $w_{\rm eff}=-\f{7}{3}$ and is a stable attractor for $\alpha_2>-2 \wedge \lambda_0>-2$, while the second one is an accelerated stable node for $\alpha_2<\lambda_0 \wedge \alpha_2<-2$ with $w_{\rm eff}<-\f{7}{3}$.
We do not consider these points viable as such values of $w_{\rm eff}$ have been already excluded by experiments (e.g.~\cite{Ade:2013zuv,Rest:2013bya}).
\item\underline{\textsl{$\phi$-MDE and $\phi$-RDE points}} \\
There are two critical points characterized by, respectively, matter and radiation domination with a non-negligible DE density:
{\small\begin{align}
& P_{5}: & & \mu_1=-\f{15}{2}, \,\, \mu_2=-3 , \,\, \mu_3=-1, \,\, \mu_4=3-\alpha_2, \,\, \mu_5=3-\lambda_0. \\
& P_{8}: & & \mu_1=-6, \,\, \mu_2=-6 , \,\, \mu_3=1, \,\, \mu_4=-\f{1}{2}-\alpha_2, \,\, \mu_5=4-\lambda_0.
\end{align}}
The first point has $\Omega_m=5$, $\Omega_{\rm DE}=-4$ and a stiff matter equation of state, while the second one has $\Omega_r=9$ and $\Omega_{\rm DE}=-8$ with $w_{\rm eff}=\f{5}{3}$. Both these points are not considered cosmologically relevant.
\item $P_9$: \underline{\textsl{unstable DE point}} 
{\small\begin{align}
&\mu_1= \alpha_2-4, \,\, \mu_2=\alpha_2-3 , \,\, \mu_3=\alpha_2-\lambda_0, \nonumber \\
& \mu_4= 6-\f{3}{4}\alpha_2-\f{1}{4}\sqrt{3}\sqrt{-\alpha_2(-32+5\alpha_2)}, \,\, \mu_5= 6-\f{3}{4}\alpha_2+\f{1}{4}\sqrt{3}\sqrt{-\alpha_2(-32+5\alpha_2)}.
\end{align}}
This point corresponds to a DE dominated configuration, albeit one that is always unstable.
\item $P_{11}$: \underline{\textsl{radiation scaling point}} \\
The stability analysis of this point is too complicated to be reported, nevertheless we are able to deduce something about its cosmological behavior. 
From Table~\ref{tabsecondorder} one can see that  the point corresponds to a scaling solution for radiation and DE with $\Omega_{DE}= 2 \alpha_2-7$. However, the constraint $\Omega_r\geq0$ imposes $\alpha_2<4$, and for this range of values the point cannot be neither a proper DE or radiation dominated point. 
\end{itemize}
\section{${\rm N}^{\rm th}$ order analysis continued}\label{apgeneral}
In this Appendix we continue with the analysis of the critical points for the $N^{\rm th}$ order system giving an overview of the points that were not presented in Sec.~\ref{recursive} since they either did not have the desired cosmological characteristics or stability.
The general structure of the critical points for the $N^{\rm th}$ order system was explained in detail in Sec.~\ref{recursive}, however here we will give a brief review. Critical points belonging to the same family can be  of  three types: ($x_c, y_c, \alpha_{0,c},\alpha_{1,c}, \alpha_{n,c}=0$) with $n\geq2$, ($x_c, y_c, \alpha_{0,c},\alpha_{1,c}, \alpha_{n,c}\neq 0)$ with $n\geq 2$ or ($x_c, y_c, \alpha_{0,c},\alpha_{1,c}, {\rm combinations} $), where `combinations' correspond to all the different combinations of $\{\alpha_{2,c},\dots,\alpha_{N-1,c}\}$ for which a different 
A thorough description of how to build all the combinations is given in Sec.~\ref{recursive}. Here we simply remind the reader that we use the index $j$ for the $\alpha_{n,c}$ in non-zero blocks that are followed by a zero-block (rule~(\ref{rule1})); while we use the index $l$ for the $\alpha_{n,c}$ of the non-zero block that closes the combination, when it exists (rule~(\ref{rule2})). Every time we substitute into~(\ref{rule1}) and~(\ref{rule2}) the specific value of $\dot{H}/H^2$ that corresponds to the point in consideration.

\begin{itemize}[leftmargin=*]
\item\underline{\textsl{Phantom DE points:}} \\
There are different families of critical points which are DE dominated but give rise to cosmological behaviors which are in tension with current observations (i.e. $w_{\rm eff}\lesssim -2$). 
However, their stable node configuration gives an attractor that, in principle, could be reached in the far future, provided that the duration of the matter era would remain long enough to allow for structure to form~(\cite{Copeland:2006wr} and references therein).
The first family that we shall consider is $P_{4a}$-like, which is a set of  DE dominated critical points with $w_{\rm eff}=-\f{7}{3}$.
\begin{subequations}
\small
\begin{eqnarray}
&&P_{4a,1}\equiv(-1, 0, 2, 0 , \alpha_{n,c}=0), \\
&&P_{4a,2}\equiv(-1,0, 2, 0, \alpha_{n,c}=\alpha_{N}+2(N-n)),\\
&&P_{4a,c}\equiv(-1, 0, 2, 0 , {\rm combinations}),\,\, \alpha_{j,c}= 2(s+1-j),\,\, \alpha_{l,c}=\alpha_{N}+2(N-l).
\end{eqnarray}
\end{subequations}
From an investigation of the eigenvalues, one finds that the first point is a stable node for $\lambda_0>-2$ while the second one exhibits this behavior for $\lambda_0>-2 \wedge \alpha_{N}<-2\l(N-2\r) \wedge \alpha_{N}>3N-8$.
The last sub-family of critical points $P_{4a,c}$ also displays stable configurations for some combinations. In that case we have $\lambda_0>-2 \wedge \alpha_{N}<-2$ and $\lambda_0>-2 \wedge \alpha_{N}>-2$. The second family that we shall consider does not  have a unique cosmological behavior, though in all the cases the critical points are DE dominated and resemble the $P_{4b}$ point of the second order analysis. 
{\small\begin{subequations}
\begin{eqnarray}
&&P_{4b,1}\equiv(-1, 0, 2, 1, \alpha_{n,c}=0),\\
&&P_{4b,2}\equiv\left(-1, 0, 2, \frac{2N-2+\alpha_{N}}{N+1}, \frac{-2n+2N+n\,\alpha_{N}}{N} \right),\\
&&P_{4b,c}\equiv\left( -1, 0, 2, \f{2s_1}{1+s_1}, {\rm combinations}\right),\,\, \alpha_{j,c}=\f{2(s+1-j)}{s+1},\,\, \alpha_{l,c}=\f{\alpha_{N}+2(N-l)+s\alpha_{N}}{s+1},
\end{eqnarray}
\end{subequations}}
where $s_1$ is the value of $s$ for the first non-zero block.
The first critical point $P_{4b,1}$ has a well defined cosmology. It is a DE dominated point with a phantom equation of state, $w_{\rm eff}=-\f{5}{3}$, and it resembles the point $P_{4b}$ of the second order with $\alpha_2=0$. 
We can infer its  stability from Table~\ref{tabsecondorder}, which shows that it is a saddle, therefore it does not have the desired nature for a DE point and we do not analyze it further.
The second critical point $P_{4b,2}$ can be written as
{\small \begin{equation}
\left(-1,0,2, 1+\f{\alpha_{2,c}}{2},\alpha_{n,c}=\f{4-2n+n\alpha_{2,c}}{2} \right),
\end{equation}}
 where we have used the solution of $\alpha_2$ to substitute for $\alpha_N$ in terms of $\alpha_{2,c}$; comparing it with Table~\ref{tabsecondorder} we can see a clear connection with the $P_{4b}$ critical point. As expected the equation of state for the effective fluid equation can be written  as
{\small \begin{equation}
w_{\rm eff}=\f{-3N-4+2\alpha_{N}}{3N}=-\f{5-\alpha_{2,c}}{3},
\end{equation}}
which is equivalent to the one found at second order for the point $P_{4b}$, and shows an accelerated behavior for $\alpha_{2,c}<4$. 
For this critical point is very difficult to calculate explicitly the eigenvalues but looking at Table~\ref{tabsecondorder} we can infer that for $\alpha_2<4$ it will be a saddle, therefore we do not consider it cosmologically viable.
In the latter case the critical points $P_{4b,c}$ has $w_{\rm eff}=-\f{7+3s}{3(s+1)}$, which for all the combinations is $\approx -1$. The stability analysis, however, reveals that this is a set of saddle points, thus preventing them from being viable accelerated attractors.

The third family of DE dominated critical points is $P_{6}$-like with $w_{\rm eff}=\f{2}{3}\lambda_0-1$:
{\small\begin{subequations}
\begin{eqnarray}
&&P_{6a}\equiv\left(\f{\lambda_0}{2},1+\f{\lambda_0}{2}, -\lambda_0, 0 , \alpha_{n,c}=0\right), \\
&&P_{6b}\equiv\left(\f{\lambda_0}{2},1+\f{\lambda_0}{2}, -\lambda_0, 0 , \alpha_{n,c}=\alpha_{N}-\lambda_0\,\l(N-n\r)\right), \\
&&P_{6c}\equiv\left(\f{\lambda_0}{2},1+\f{\lambda_0}{2}, -\lambda_0, 0 , {\rm combinations}\right),\,\, \alpha_{j,c}= -(s+1-j)(3+\lambda_0),\,\, \alpha_{l,c}=\alpha_{N}-(N-l)(3+\lambda_0).
\end{eqnarray}
\end{subequations}}
The eigenvalues of the linearized system around these critical points are too complicated to be reported.
However it can be shown that the first one is an accelerated attractor for $(-\f{12}{5}<\alpha_{N}\leq-2 \wedge -\f{12}{5}\leq\lambda_0<\alpha_{N}) \vee (\alpha_{N}>-2 \wedge -\f{12}{5}\leq\lambda_0<-2)$ while the second one displays the same cosmological behavior for $(\alpha_{N} < \f{1}{5} (24 - 12 N) \wedge -\f{12}{5} \leq \lambda_0 < -2) \vee (\alpha_{N} =\f{1}{5} (24 - 12 N) \wedge -\f{12}{5} < \lambda_0 < -2) \vee ( \f{1}{5} (24 - 12N) < \alpha_{N} < 4 - 2N \wedge \f{\alpha_{N}}{-2 + N} < \lambda_0 < -2)$. 
Both these points, as well as $P_{6c}$ have $w_{\rm eff}<-2.3$, therefore we do not consider them cosmologically viable.

The last family of, $P_9$-like, DE dominated critical points contains configurations which all have a different effective equation of state.
{\small\begin{subequations}
\begin{align}
P_{9a} \equiv & (-5, 0, 6, 3, \alpha_{n,c}=0),\\
P_{9b}\equiv & \left(\frac{-1-2N+\alpha_{N}}{\l(N-1\r)},  0, \frac{3N-\alpha_{N}}{\l(N-1\r)}, 3, \frac{3N+n(\alpha_{N}-3)-\alpha_{N}}{\l(N-1\r)}\right) \nonumber \\
 =&  \left(\alpha_2-5, 0, 6-\alpha_2, 3, \alpha_{n,c}=6+n(\alpha_2-3)-\alpha_2 \right),\\
P_{9c}\equiv & \left(-2-\f{3}{s_1}, 0, 3+\f{3}{s_1}, 3, {\rm combinations} \right),\,\, \alpha_{j,c}= \f{3(s+1-j)}{s},\,\, \alpha_{l,c}=\f{3(N-l)+s\alpha_{N}}{s},
\end{align}
\end{subequations}}
where $s_1$ is the value of $s$ for the first non-zero block. The equation of state parameter in these three configurations is, respectively:
{\small \begin{align}
w_{\rm eff}(P_{9a})=-3 ,\,\,  w_{\rm eff}(P_{9b})= -\f{3+3N-2\alpha_{N}}{3\l(N-1\r)}=-3+\f{2}{3}\alpha_2 ,\,\, w_{\rm eff}(P_{9c})=-\f{2+s_1}{s_1}.
\end{align}}
The stability analysis reveals that all the points of this family are saddles in the range for which they are accelerated and therefore we do not investigate them further.
\item\underline{\textsl{Scaling solutions:}}\\
This family of critical points is characterized by a scaling between matter and DE:
{\small\begin{subequations}
\begin{align}
P_{sc1} \equiv & \bigg( \frac{-3-6\l(N-1\r)^2+5\alpha_{N}-2\alpha_{N}^2+\l(N-1\r)(-9+7\alpha_{N})}{3(N-2)^2}, 0, \frac{3N-2\alpha_{N}}{N-2}, \frac{3\l(N-1\r)-\alpha_{N}}{N-2},  \nonumber \\
& \hspace{10cm} ,\alpha_{n,c}=\frac{3N+n(-3+\alpha_{N})-2\alpha_{N}}{N-2}  \bigg), \\
P_{sc2} \equiv & \left(-\f{(s_1+1)(2s_1+1)}{(s_1-1)^2}, 0, \f{3(s_1+1)}{s_1-1}, \f{3s_1}{s_1-1}, {\rm combinations} \right),\,\, \alpha_{j,c}=\f{3(s+1-j)}{s-1},\,\, \alpha_{l,c}=\f{-\alpha_{N}+3(N-l)+s\alpha_{N}}{s-1},
\end{align}
\end{subequations}}
where $s_1$ is the value of $s$ for the first non-zero block. These configurations correspond to a matter density and equation of state parameter:
{\small \begin{align}
\Omega_m(P_{sc1}) =-\f{(4 +N  - 2 \alpha_{N}) (-3 + \alpha_{N})}{3 ( N-2)^2} ,\,\, w_{\rm eff}(P_{sc1})=\f{ 3 N - 2 \alpha_{N}}{6- 3 N},\,\, \Omega_m(P_{sc2})=\f{5+s_1}{(s_1-1)^2} ,\,\,  w_{\rm eff}(P_{sc2})=\f{s_1+1}{1-s_1}.
\end{align}}
The study of the stability for these critical points is very complicated due to the unknown value of N. It is, however, simple to determine that, for both points, neither of the two configurations in which they are, respectively, matter ($\Omega_m=1$) and DE dominated ($\Omega_{\rm DE}=1$) is cosmologically viable.

In this paper we choose not to perform a full analysis of the scaling configurations, but rather focus on the two extrema for which either of the two components has fractional energy density equal to unity. While we leave a thorough investigation of the scaling regime for future work, we want to stress that this family of critical points is expected to display all the late-time scaling cosmologies that can offer a dynamical solution to the coincidence problem~\cite{Copeland:1997et,Tsujikawa:2004dp,Gomes:2013ema}.

\item\underline{\textsl{DE points:}} \\
The last family of critical points is made of DE dominated configurations
{\small \begin{subequations}\begin{align}
P_{d1} \equiv & \bigg( \frac{-2\alpha_{N}^2-\lambda_0(-3+\l(N-1\r)^2(\lambda_0+1)+\l(N-1\r)(\lambda_0+2))+\alpha_{N}(-4+\lambda_0+\l(N-1\r)(4+3\lambda_0))}{6(N-2)^2}, \nonumber\\ 
& \hspace{0.5cm} ,\frac{(\alpha_{N}-3)(-2+2\alpha_{N}-\lambda_0)+\l(N-1\r)^2(6-5\lambda_0+\lambda_0^2)+\l(N-1\r)(-12+\alpha_{N}(8-3\lambda_0)+2\lambda_0+\lambda_0^2)}{6(N-2)^2},\nonumber\\ 
& \hspace{4.3cm} ,\frac{\lambda_0+\l(N-1\r)\lambda_0-2\alpha_{N}}{N-2}, \frac{\l(N-1\r)\lambda_0-\alpha_{N}}{N-2}, \alpha_{n,c}=\frac{(n-2)\alpha_{N}+(N-n)\lambda_0}{N-2}\bigg), \\
P_{d2} \equiv & \bigg( -\f{\lambda_0(-3+s_1^2(\lambda_0+1)+s_1(\lambda_0+2))}{6(s_1-1)^2},  \f{3(2+\lambda_0)+s_1^2(6-5\lambda_0+\lambda_0^2)+s_1(-12+2\lambda_0+\lambda_0^2)}{6(s_1-1)^2}, \nonumber\\ 
& \hspace{2.8cm} ,\f{\lambda_0(s_1+1)}{s_1-1},\f{s_1\lambda_0}{s_1-1}, {\rm combinations} \bigg),\,\, \alpha_{j,c}=\f{(s+1-j)\lambda_0}{s-1},\,\, \alpha_{l,c}=\f{\alpha_{N}(s-1)+(N-l)\lambda_0}{s-1},
\end{align}\end{subequations}}
 with different values of the equation of state, respectively
{\small\begin{align}
w_{\rm eff}(P_{d1})=\f{6-3N+2\alpha_{N}-2\lambda_0}{3(N-2)} \,,\,\,\,\,\,\, w_{\rm eff}(P_{d2})=\f{3-3s_1-2\lambda_0}{3(s_1-1)},
\end{align}}
where $s_1$ is the value of $s$ for the first non-zero block. The first point has a viable cosmological behavior for $\alpha_{N}<N-2+\lambda_0$  and would have $w_{\rm eff}=-1$ if 
$\lambda_0=\alpha_{N}$; however we are not able to analyze its stability. 
The second point gives a viable cosmological behavior for $s_1+\lambda_0>1$; however requiring $w_{\rm eff}=-1$ gives $\lambda_0=0$ and the stability analysis reveals that the point is non-hyperbolic for such a value. 
\item\underline{\textsl{$\phi$-MDE:}}\\
This family contains the following $P_5$-like:
{\small\begin{subequations}
\begin{align}
& P_{5a}\equiv(-1, 0, -3, 0 , \alpha_{n,c}=0),  \\
& P_{5b}\equiv(-1, 0, -3, 0 , \alpha_{n,c}=\alpha_{N}-3(N-n)), \\
& P_{5c}\equiv(-1, 0, -3, 0, {\rm combinations}),\,\, \alpha_{j,c}= -3(s+1-j) ,\,\, \alpha_{l,c}=\alpha_{N}-3(N-l),
\end{align}
\end{subequations}}
which are characterized by $\Omega_m=5$,  $\Omega_{\rm DE}=-4$ and $w_{\rm eff}=1$, therefore we do not consider them further.
\end{itemize}


\end{document}